\documentclass[12pt]{article}
\usepackage{arxiv}
\usepackage{mathtools,amsfonts,amsmath}
\usepackage{graphicx}
\usepackage{subcaption}
\newcommand{\bmhead}[1]{\paragraph*{\textbf{#1}}}
\usepackage{hyperref}
\usepackage{multirow,multicol,colortbl,xspace,booktabs,longtable}
\newcolumntype{C}{>{\raggedleft\arraybackslash}p{3mm}}
\usepackage{float}  
\usepackage[percent]{overpic}
\usepackage{makecell}

\newcommand{\methods}{\hyperref[sec:methods]{Methods}\xspace}

\usepackage[style=numeric,sorting=none,backend=biber]{biblatex}
\date{}

\makeatletter
\renewcommand{\fnum@figure}{\textbf{Figure \thefigure}}
\renewcommand{\fnum@table}{\textbf{Table \thetable}}

\title{The Retraction Epidemic in Science Across\\Publishers, Fields, and Countries}

\author{
Sara Venturini \\
MoBS Lab\\ 
Northeastern University\\
Boston 02115, MA, US\\
\And
Alessandra Urbinati \\
MoBS Lab\\ 
Northeastern University\\
Boston 02115, MA, US\\
\And
Paola Gallo \\
Dipartimento di Matematica e Fisica\\ 
Università Roma Tre\\
Rome 00154, Italy\\
\And
Jessica T. Davis \\
MoBS Lab\\ 
Northeastern University\\
Boston 02115, MA, US\\
\And
Alessandro Vespignani \\
MoBS Lab\\ 
Northeastern University\\
Boston 02115, MA, US\\
Institute for Scientific Interchange Foundation\\ 
Turin 10123, Italy\\
}

\begin{document} 

\maketitle

\begin{abstract} 
Retractions serve as an indicator of failures in research integrity, yet most analyses focus on absolute counts rather than risk per paper. We use one of the largest open bibliographic databases to develop incidence metrics normalized by population: retractions per publication and per active author annually. Applying an epidemiological framework that models counts with exposure, we find evidence of exponential growth in retraction incidence, with approximately a 5-year doubling time at both the paper and author levels. These patterns vary significantly across fields, publishers, and countries. While scientific output is becoming more democratized globally, retractions are concentrated in fewer countries, creating a “concentration” paradox that calls for targeted monitoring. Despite exponential growth, the absolute incidence remains low ($0.12\%$ in 2021), allowing for corrective intervention. Incidence-based monitoring provides a framework for evaluating policies that safeguard research integrity at scale.
\end{abstract} 

\newpage
\section{Introduction}\label{sec:introduction}
\noindent The phenomenon of scholarly article retractions has gained attention due to its essential role in maintaining scientific integrity and correcting the published record~\cite{brainard2018massive,fang2012misconduct,van2011science,van2023news,van2023big}. Unfortunately, the past two decades have seen a sharp rise in the absolute number of retractions, culminating in $2023$ with a new annual record of over $10,000$ retracted articles~\cite{van2023news}. This surge raises a fundamental question: does the rise in retractions reflect a genuine deterioration in research integrity, improved detection of problematic work, or simply the natural consequence of exponential growth in scientific output?
Retractions, although imperfect, serve as an operational proxy for research-integrity failures (combining misconduct and serious error). Prior large-scale studies estimate that misconduct accounts for a majority of retractions in biomedical domains~\cite{fang2012misconduct,freijedo2024biomedical,marco2021fraud, song2025systematic, gaudino2021trends}, but practices vary across fields; accordingly, we focus on the incidence of retractions as an indicator of integrity stress rather than equating retractions with misconduct per se. Moreover, retracted articles continue to accumulate citations and generate second-order citations, propagating unreliable findings~\cite{bar2018temporal, ioannidis2025linking,wray2018retractions,stavale2019research,candal2022retracted,candal2024retracted, shepperd2023analysis}. Understanding whether retraction rates are accelerating is thus critical for safeguarding the reliability of the scientific enterprise.

Bibliometric methodologies analyzing data aggregated from major databases are the common approach for characterizing retraction trends. However, many analyses rely on raw retraction counts, which can be misleading in the context of a rapidly expanding publication output; increases in absolute retractions may reflect the increase in total publication rather than a genuine increase in misconduct or error.~\cite{wray2018retractions}. This methodological limitation mirrors a classic epidemiological problem: observing more cases of a disease does not necessarily mean the disease is becoming more prevalent if the population at risk is also growing. What remains insufficiently and unevenly addressed is the systematic, dynamic characterization of incidence-based metrics, such as new retractions per unit of output, across fields, publishers, and geographic regions~\cite{fang2012misconduct,van2023big,gaudino2021trends,stavale2019research,candal2022retracted,shepperd2023analysis,Steen249,chen2013visual,grieneisen2012comprehensive}.

In this study, we use an epidemiological framework to define the per-paper and per-author retraction risk over time. By modeling retraction counts as a function of exposure, we can statistically test whether retraction risk remains constant, increases linearly, or grows exponentially. Our approach identifies the overall incidence trajectory of retractions and identifies substantial heterogeneity across scientific disciplines, publishing organizations, and national research systems. These findings have direct implications for resource allocation, monitoring strategies, and the design of interventions to maintain trust in science.
\\

\section{Results} \label{sec:results}

\begin{figure}[ht!]
\centering
\includegraphics[width=\textwidth]{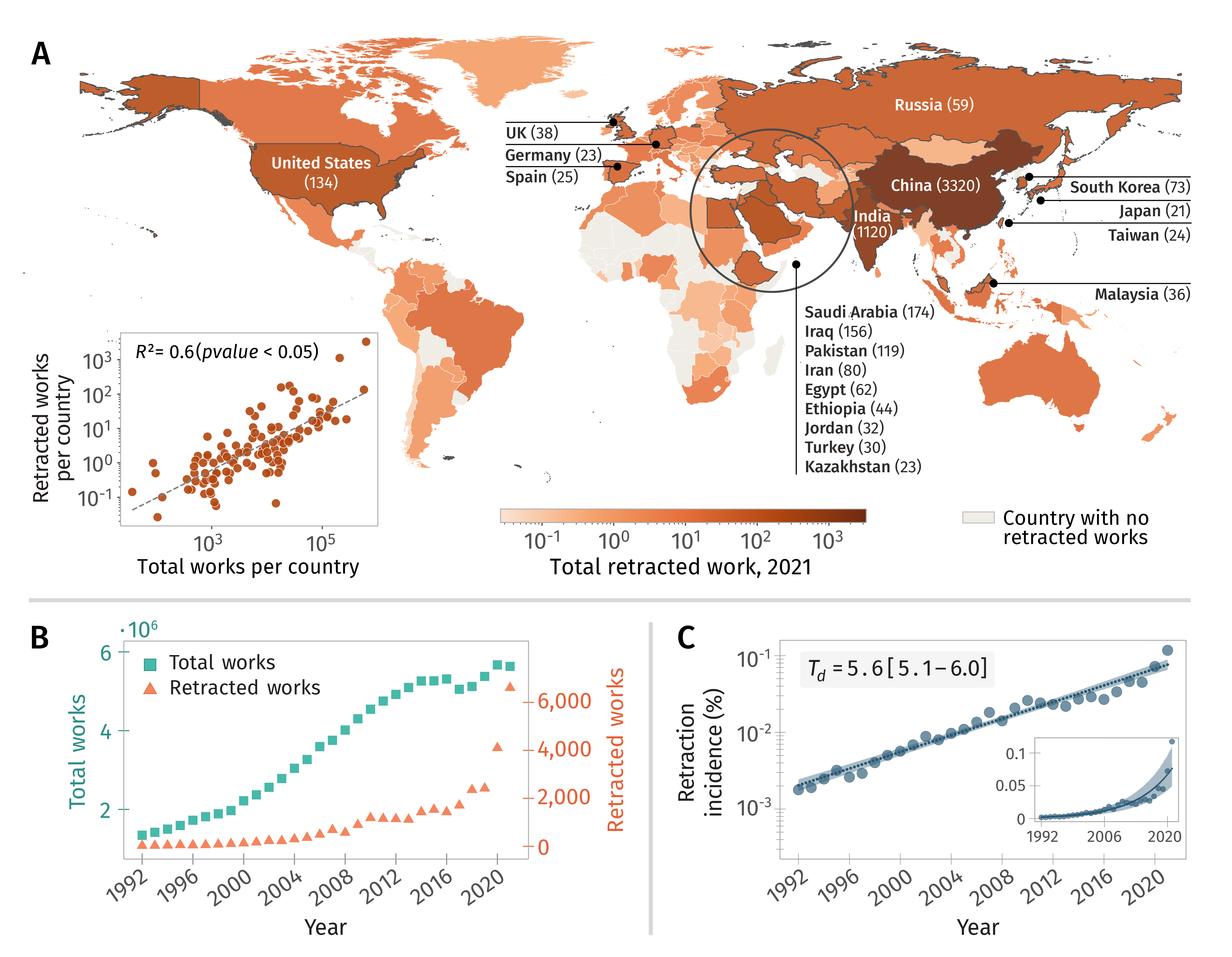}
\caption{\textbf{Global and temporal patterns of scientific retractions.}
\textbf{A.} World map of raw retraction counts by country in 2021 (log color scale). Countries with zero retractions (or missing data) are shown in grey. Selected countries are annotated with rounded retraction counts. Inset: Pearson correlation between total publications and retractions by country in 2021, restricted to countries with at least one retraction ($R^2=0.60$, $P\leq 0.05$).
\textbf{B.} Annual totals of published works ($N_t$, green squares) and retracted works ($R_t$, orange triangles), 1992--2021. Both series rise over time, illustrating why raw counts alone cannot diagnose changes in per-paper risk.
\textbf{C.} Annual retraction \emph{incidence} (blue circles; percent of works published in year $t$ that are eventually retracted), with an exposure-adjusted GLM fit assuming exponential growth and a negative binomial likelihood (blue line; growth rate significant at $P\leq 10^{-4}$). Shaded band shows the 95\% confidence interval for the mean incidence. The legend reports the estimated doubling time $T_d$ with its 95\% CI. Incidence is plotted on a logarithmic $y$-axis. The inset report the linear-scale plot with the 95\% prediction interval.
}
\label{fig:works}
\end{figure}
\subsection{Data Sources and Incidence Metrics}
Here, we analyze published works in the OpenAlex dataset~\cite{priem2022openalex} (October $2025$ updated snapshot). OpenAlex is one of the largest open bibliometric dataset, and we restrict the analysis to records with a known publication year, article type, and sources classified as journal or conference proceedings. This results in over $107$M works published from $1992$ to $2021$, of which about $31,000$ are flagged as retracted in OpenAlex  through a boolean entry sourced from the Retraction Watch database~\cite{retractionwatch2023}. OpenAlex also provides additional metadata such as active-author counts, topic tags mapped to fields and domains, institutional affiliations, and publisher identifiers, enabling multi-level analyses across disciplines, countries, and publishers. For each publication year $t$, we record the total number of works published in that year and the subset that were retracted at any point in time (i.e., counts are indexed by publication cohort). This is an important point as we consider the incidence per year of publication and not retraction date. Because retractions typically occur with a delay after publication, we analyze the delay distribution for works published since $1992$; in the last decade we observe a median lag of approximately $2.0$ years ($95\%$ confidence interval (CI) $[1.0\text{–}3.0]$) (Appendix Fig.~\ref{fig:time to retraction}), consistent with previous findings~\cite{steen2016has}. To prevent right-censoring in recent cohorts, we limit our analysis to papers published before $2021$, ensuring that retraction counts are sufficiently stable for the incidence analyses that follow (see \methods section and the Appendix for additional details). 

\subsection{Exponential Growth of Retraction Incidence} 
Fig.~\ref{fig:works}A maps retraction counts by country in 2021. These counts are strongly associated with each country’s publication volume ($R^2\simeq 0.6$), meaning places that publish more papers also register more retractions. In other words, higher retraction counts can arise mechanically from a larger base of published works.
A similar confounding appears over time. In Fig.~\ref{fig:works}B we plot, for each year $t$ from $1992$ to $2021$ (a $30$-year window), the total number of published works $N_t$ and the number of retracted works $R_t$. Both series grow over time, and it is not possible to conclude whether the increase in retractions simply mirrors the expansion of overall output or reflects an acceleration in the \emph{per-paper} risk of retraction. To distinguish these mechanisms, we measure the annual rate of retracted works defined as $R_t/N_{t}$.
In Fig.~\ref{fig:works}C we show the approximate linearity of the \emph{log}-incidence as a function of time. This behavior suggests an exponential increase in retraction incidence. To provide a principled statistical analysis, we assume that each year $t$ generates $N_t$ “opportunities” for retraction (the number of papers published that year). The observed retractions are then realizations of an exposure counting process with incidence $i_t$; under independent, rare events this yields the Poisson model with 
\begin{equation}
i_t=\frac{\mathbb{E}[R_t]}{N_t}, 
\end{equation}
where $\mathbb{E}[R_t]$ is the expected value of the retraction count at time $t$.
We consider three ways the per-paper retraction risk (incidence) may evolve over time: \emph{constant incidence}, \emph{linearly changing incidence}, and \emph{exponential-growth incidence}. We fit according to each specification the signal $R_t$, performing a regression analysis with generalized linear models (GLMs)~\cite{nelder1972generalized} with an \emph{offset} $\log N_t$. This offset accounts for the size of the publication base in that year. As shown in the Appendix, mean--variance diagnostics indicate clear overdispersion relative to a Poisson process. We therefore estimate negative binomial (NB) GLMs with the same link and exposure offset, allowing the dispersion  parameter to be estimated from the data rather than fixed a priori. Full model details and robustness checks are reported in the \methods section.
 Using a BIC-based model selection procedure, the evidence overwhelmingly favors the exponential-growth incidence specification, which implies
\begin{equation}
i_t\;=\; i_0\, e^{\,g(t-t_0)},
\end{equation}
 where $g$ is the incidence per-year growth rate and $i_0$ is the baseline incidence at time $t_0$. Full model selection details and robustness checks are reported in the \methods section. An exponential increase in incidence implies  that the per-year growth rate $g$ can be associated to the \emph{doubling time}, i.e. the number of years required for incidence to double, through the relation 
$T_d =ln2/g$. Because $g$ is estimated from data, it has an associated uncertainty that propagates to $T_d$, producing confidence intervals for the doubling time.
Using the estimated growth rate over the last $30$ years, we obtain a doubling time for the retracted-works incidence of $5.6$ years (95\% CI $[5.1-6.0]$). This indicates that the per-paper risk of retraction doubles approximately every five and a half years within the study window.  This multiplicative dynamic accelerates the  straining of editorial capacity in dealing with retractions, and has the potential of quickly eroding trust in the scientific publishing system.

 \subsection{Author-Level Incidence}
To determine whether exponential growth reflects phenomena at the paper level or broader systemic patterns, we analyze retraction rates at the author level. For each year $t$, we define author exposure $A_t$ as the number of active authors (those publishing at least one work in year $t$), with the outcome being the number of authors with one or more retractions from works published that year. Using the same exposure-adjusted GLM framework, we find that author-level incidence also shows exponential growth. The estimated doubling time is 5.7 years (95\% CI [5.1-6.4]), closely matching the paper-level estimate of 5.6 years (Appendix Fig.~\ref{fig:authors}, Tables~\ref{tbl:authors_ranges}--~\ref{tbl:authors}).
The consistent behavior of authors and retractions incidences is consistent across various time windows, with author-level doubling times ranging from 5.2 to 8.3 years (Appendix Table~\ref{tbl:authors}). This convergence strongly supports the presence of a system-wide multiplicative process. The observed coupling suggests that retraction risk scales with both research output and the number of active researchers, pointing to systemic pressures (publication metrics, career incentives) that shape research practice broadly rather than affecting only isolated individuals.

\subsection{Domain and field heterogeneity} 
\begin{figure}
\centering
\includegraphics [width=0.80\textwidth, keepaspectratio]{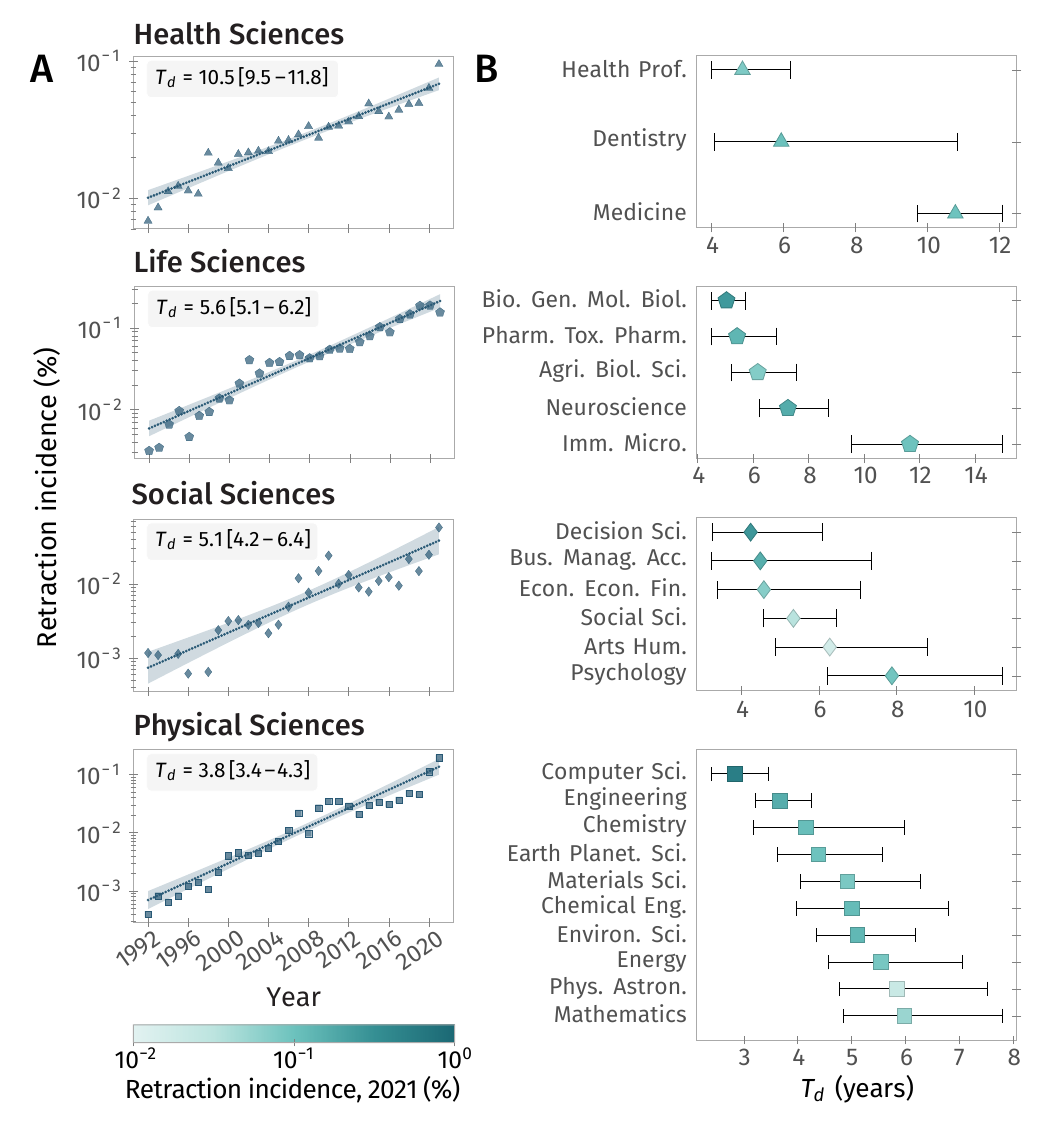} 
\caption{
\textbf{Retraction incidence over time across domains and doubling time by field.}
\textbf{A.} Annual retraction \emph{incidence} (blue points) by \emph{domain}, fit with an exposure-adjusted GLM assuming exponential growth and a negative binomial likelihood (blue line). Shaded bands denote 95\% confidence intervals on the mean values ($P$-values and statistical analysis are reported in the Appendix). The legend reports the estimated doubling time $T_d$ with its 95\% confidence intervals. Incidence is shown on a logarithmic $y$-axis
\textbf{B.} Estimated doubling times ($T_d$) for the incidence of retracted works by \emph{field} (nested within domains) with strong statistical support for exponential growth. Points show $T_d$ estimates; horizontal bars are 95\% confidence intervals. Marker color encodes the field’s retraction incidence percentage in $2021$ (log color scale; see color bar).
See Table~\ref{tbl:abbreviations} for field abbreviations.}
\label{fig:domains}
\end{figure}
Prior work indicates that retraction trends can vary across domains and fields, reflecting differences in publication volume, scrutiny intensity, and integrity issues of each discipline~\cite{gaudino2021trends,shepperd2023analysis,grieneisen2012comprehensive}. To characterize this heterogeneity across domains and fields, we map each work to a specific domain/field using the OpenAlex topic hierarchy and apply fractional weighting to avoid double counting (see Appendix). For every stratum and year, we recompute the weighted publication total $N_t$
and retraction total $R_t$ and we estimate incidence trends with the same exposure-adjusted GLM framework used at the aggregate level. This yields directly comparable growth estimates (and doubling times) across domains and fields.

For both categories, the incidence is well described by an exponential increase over the study window. However, the growth rate and doubling time vary substantially by area. Fig.~\ref{fig:domains}A shows representative exponential fits at the domain level, highlighting good log-incidence linearity and stable growth estimates.
Fig.~\ref{fig:domains}B shows a clear spread in doubling times across fields: Computer Science appears shortest, followed closely by Engineering, with similar values for Chemistry and Earth \& Planetary Sciences. These areas have relatively small confidence intervals, reinforcing the signal for the multiplicative increases of incidence. 
Much of the physical sciences (Materials Science, Chemical Engineering, Energy, Physics \& Astronomy, Environmental Science, Mathematics), several life sciences (Pharmacology/Toxicology, Biochemistry/Genetics, Agricultural \& Biological Sciences, Neuroscience), and many social science areas (Decision Sciences, Economics/Finance, Business/Management, Social Sciences, Arts \& Humanities) shows doubling times in the range of $4\text{–}7$ years. Within the health sciences, the Nursing and Veterinary subfields have limited sample sizes. In these cases, the evidence for an exponential incidence is not definitive, and alternative trend forms cannot be ruled out.
Interestingly, we also observe that volume and growth are not the same signal: Medicine shows high retraction volume but not the fastest growth, whereas some fast-growing fields have lower volumes.



\subsection{Regional Patterns and Publisher Effects} 
\begin{figure}
\centering
\includegraphics[width=\textwidth] {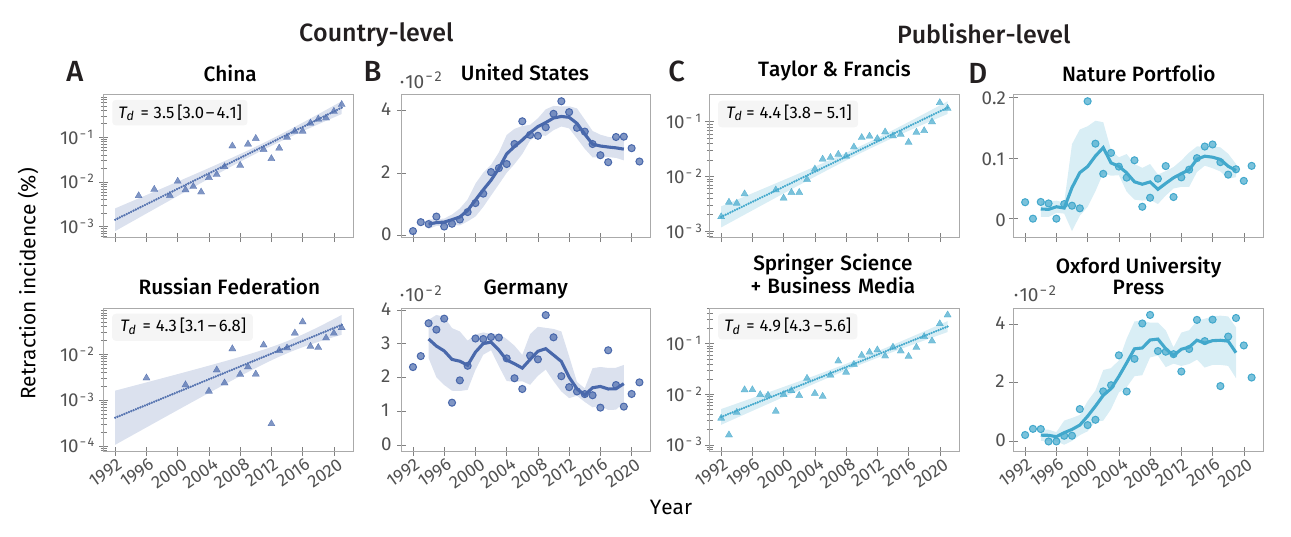} 
\caption{\textbf{Retraction incidence over time across countries and publishers.}
\textbf{A.} Country-level retraction \emph{incidence} for cases with strong statistical support for exponential growth (blue triangles), with a negative binomial, exposure-adjusted exponential fit (blue line; growth-rate $P\leq 0.05$). Shaded band: 95\% CI of the mean values. The legend reports the estimated doubling time $T_d$ (95\% CI). Logarithmic $y$-axis.
\textbf{B.} Country-level retraction \emph{incidence} (blue circles) for cases without strong support for any candidate model, shown with a centered 5-year rolling average (blue line) and $\pm$1~SD band (shaded).
\textbf{C.} Publisher-level retraction \emph{incidence} for cases with strong statistical support for exponential growth (cyan triangles), with a negative binomial, exposure-adjusted exponential fit (cyan line; growth-rate $P\leq 0.05$). Shaded band: 95\% CI of the mean values. The legend reports $T_d$ (95\% CI). Logarithmic $y$-axis.
\textbf{D.} Publisher-level retraction \emph{incidence} (cyan circles) for cases without strong model support, shown with a centered 5-year rolling average (cyan line) and $\pm$1~SD band (shaded).
 }
\label{fig:countries_publishers}
\end{figure}

Geographical variations in retractions reflect differing national research landscapes~\cite{brainard2018massive,KOO2024e38620}.
China and the United States account for the highest absolute number of retractions ($2003$–$2022$ data) due to their massive scientific output. However, it has been observed that when accounting for publication volume, countries with smaller scientific output often show the highest retraction incidence values~\cite{freijedo2024biomedical,KOO2024e38620}.
We quantify country-level retraction incidence by mapping papers to institutions and institutions to countries, then allocating each paper’s unit weight proportionally across co-authors and take into account multi-affiliations (see Appendix). For each country and year, we recompute the weighted publication total 
$N_t$ and the weighted retraction total 
$R_t$, and estimate trends with the same exposure-adjusted GLM framework used previously. It is worth noting that the statistical  signal varies with the output volume. For countries with limited publication weight and few retractions, the results tend to be noisier and should be interpreted with caution.
Within these limits, clear patterns emerge. 
A few countries exhibit a sustained exponential increase in retraction incidence robust to sensitivity analysis in the time window considered (see Appendix). Among those China ($T_d = 3.5$ years, $95\%$ CI $[3.0\text{–}4.1]$), and the Russian Federation ($T_d = 4.3$ years, $95\%$ CI $[3.1\text{–}6.8]$), see Fig.~\ref{fig:countries_publishers}A.
Other countries show different behaviors. The United States, for instance, shows signs of course correction: after a period of earlier growth, recent trajectories are flat or declining, which could potentially indicate stronger screening or editorial policies (Fig.~\ref{fig:countries_publishers}B). Germany stands out over the full window, with an overall reduction in incidence across three decades (Fig.~\ref{fig:countries_publishers}B). These patterns suggest that, while exponential growth persists in some countries, policy and workflow reforms can successfully bend the curve. 

Similar hetereogenous results are observed by analyzing retraction incidence at the publisher level. We assign each work to its publisher using OpenAlex metadata (journal–to–publisher mapping). The statistical signal is significative for large publishing houses; for smaller publishers with limited output, estimates carry wider uncertainty and shall be interpreted cautiously.
Several major groups exhibit a clear exponential increase in retraction incidence over the study window. In particular, Taylor \& Francis and Springer Science \& Business Media show multiplicative dynamics with estimated doubling times in the ranges $4.4$ years $(95\%$ CI $[3.8\text{–}5.1])$, and $4.9$ years $(95\%$ CI $[4.3\text{–}5.6])$, respectively (see Fig.~\ref{fig:countries_publishers}C). By contrast, some publishers now show flat or declining incidence: e.g., Nature Portfolio, and Oxford University Press, although we cannot conclude if this is due to tighter screening and quality controls (Fig.~\ref{fig:countries_publishers}D). Overall, trends are heterogeneous: some houses are on a corrective path while others sustain exponential growth, suggesting that incidence-based monitoring and targeted policy evaluation at the publisher scale may be useful in contrasting the growth of retractions.

\subsection{Relative Retraction Incidence}
Despite the exponential increase, the total incidence of retraction remains low, at approximately $0.12\%$ during our study period. 
\begin{figure}[ht!]
\includegraphics[width=\textwidth]{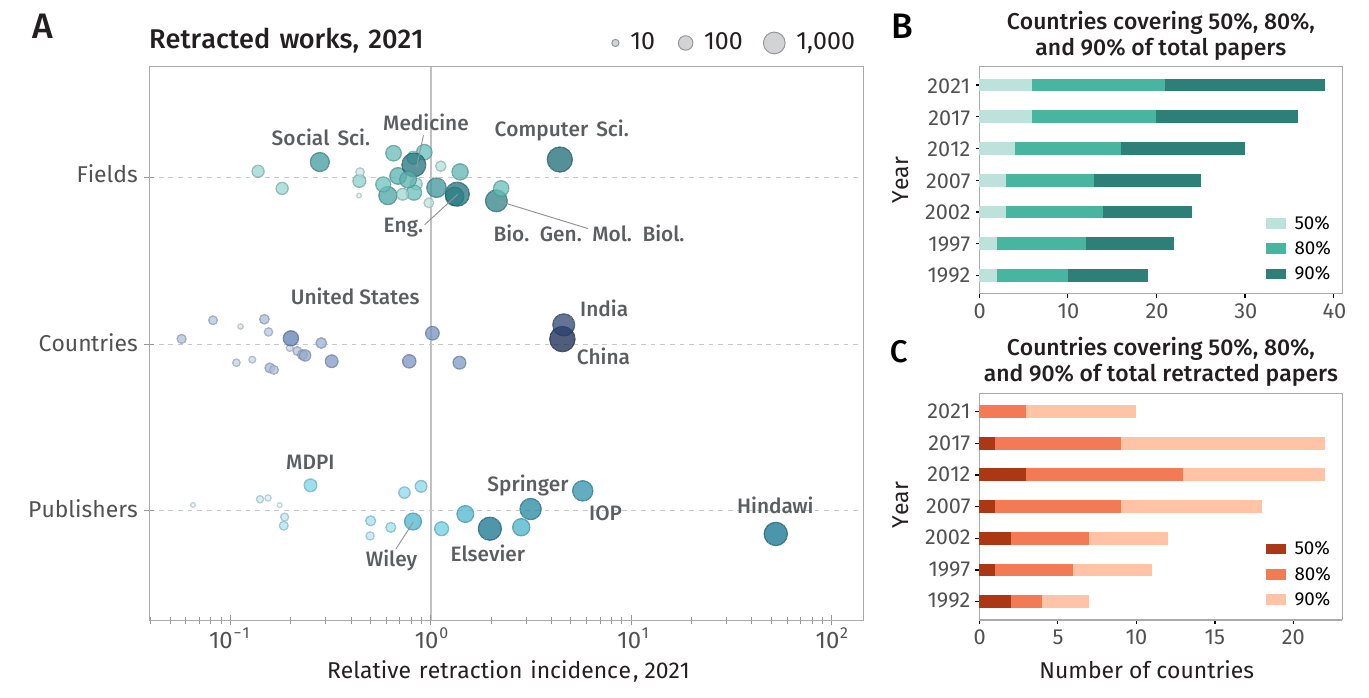}
\caption{\textbf{Relative retraction incidence and concentration paradox.}
\textbf{A.} Relative retraction \emph{incidence} (see definiton in the text) across \emph{fields}, \emph{countries}, and \emph{publishers} in 2021. Each point represents one entity; its position on the $x$-axis is the entity’s relative incidence (log scale). Point size and color intensity encode the total number of retracted works in 2021 (log scale). Entities are grouped by type with dashed horizontal separators; representative labels mark discussed cases. See Table~\ref{tbl:abbreviations} for field and publisher abbreviations. Number of countries required to account for 50\%, 80\%, and 90\% of global output at five-year intervals (1992--2021), illustrating the concentration paradox. {\bf B:} total publications—increasing country counts indicate democratization. {\bf C:} total retractions—decreasing country counts indicate concentration in fewer countries. This divergence is reinforced by Gini coefficient analysis (Appendix Fig.~\ref{fig:Gini_countries}): Gini for publications declines from 0.92 to 0.88, while Gini for retractions increases from 0.55 to 0.90, with even more extreme publisher-level concentration (Gini 0.45 to 0.90, Appendix Fig.~\ref{fig:Gini_publishers}).
}
\label{fig:jitter}
\end{figure}
In Fig.~\ref{fig:jitter}A we show the relative retraction incidence across research fields, countries, and publishers. The relative retraction incidence is defined as the ratio between the retraction incidence of a given field, country, or publisher and the overall retraction incidence in the same year. Thus, a relative incidence of $1$ indicates a retraction rate equal to the overall publication rate, while a value greater than $1$ denote proportionally higher rates, that is, retractions occurring at a rate multiplied by that factor relative to the global trend. The Fig.~\ref{fig:jitter}A indicates substantial heterogeneity across organizational levels. 
At the field level, most fields cluster around or below the global baseline, while Computer Science stands out with both elevated relative incidence (around $4.5$ times the global rate) and a high absolute numbers of retracted works. This result should be interpreted with caution, however, as conference proceedings, the primary publication venue in Computer Science, are indexed inconsistently across bibliographic databases, and their variable coverage may inflate apparent retraction rates relative to fields that publish predominantly in journals.
At the country level, China and India, the two countries with the largest absolute numbers of retracted works, exhibit relative incidence rates of approximately $4.5$ times the global average, indicating that their high retraction counts reflect both high publication volumes and elevated per-capita rates. Most other countries cluster near or below the global baseline. While the United States ranks third in absolute numbers of retracted works, its relative incidence rate is only about $0.2$.
At the publisher level, most publishers show relative incidence below $10$. The Hindawi outlier publisher~\cite{van2023news} also suggests that policies and workflows can substantially alter incidence, supporting the case for incidence-based monitoring (with lag-aware reporting) and publisher-specific remediation rather than one-size-fits-all solutions.

\subsection{The Concentration Paradox}
An important observation emerges from examining how retractions are distributed geographically and institutionally over time. Scientific output is becoming more globally distributed. The number of countries accounting for 90\% of total publications increases from about 25 in 1992 to nearly 40 by 2021 (Fig.~\ref{fig:jitter}B). Retractions follow the opposite trend: the number of countries accounting for 90\% of total retractions decreases from roughly 15 to fewer than 10 over the same period (Fig.~\ref{fig:jitter}C). Gini coefficient analysis confirms this divergence (Appendix Fig.~\ref{fig:Gini_countries}): the Gini for publications declines from 0.92 to 0.88, while the Gini for retractions rises from 0.55 to 0.90. Publisher-level concentration is even more pronounced (Appendix Figs.~\ref{fig:Coverage_publishers}--~\ref{fig:Gini_publishers}): just 2--3 publishers account for 50\% of all retractions in 2021, with the retraction Gini rising from 0.45 to 0.90 over three decades.





\section{Discussion}\label{sec:discussion}
The convergence of evidence across multiple levels of analysis indicates a system-wide multiplicative process driving retraction rates. At the overall level, log-incidence increases nearly linearly over three decades, showing a pattern of exponential rather than additive growth. This trend is consistent across various aggregation levels: exposure-adjusted GLM fits reliably favor the exponential-growth model across different fields, publishers, and countries, ruling out explanations based on a few outlier sectors.  

The author-level analysis provides additional support to this interpretation. The same trend at both the output and contributor levels is difficult to attribute solely to editorial factors. It is consistent with mechanisms that propagate risk across researchers, such as outsourced authorship pipelines or norm diffusion through collaborative networks. This pattern can be contextualized within the framework of alternative measures of research integrity. For instance, self-report surveys~\cite{fanelli2009many} suggest approximately 2\% of scientists admit to serious misconduct, likely an underestimate given reporting bias. In addition, studies of predatory publishing find that more than half of authors in such venues publish there only once~\cite{steen2016has,frandsen2022authors}, consistent with a diffuse, system-level stress rather than a narrow misconduct problem. These observations imply that interventions must address both editorial practices and the broader incentive structures that influence researcher behavior~\cite{azoulay2017career,azoulay2015retractions,memon2025characterizing}.

A purely additive mechanism cannot explain the above patterns. The per-paper risk multiplies by a roughly constant factor each year. This has a direct practical consequence. Small reductions in the growth rate $g$, through tightened screening, provenance checks, paper-mill detection, or audit triggers~\cite{van2023big,abalkina2025stamp}, bend the curve disproportionately over time. The reverse is equally true: delayed intervention lets multiplicative effects work against intervention policies, steadily raising the effort required to stabilize incidence.

The low absolute incidence (0.12\% in 2021) means corrective action remains feasible. The heterogeneity in relative incidence across fields, countries, and publishers, however, implies that uniform global policies will miss the most affected segments. The Hindawi case is instructive: a single publisher's can substantially alter an incidence trajectory, demonstrating that publisher-specific policies are both possible and effective~\cite{van2023news}.
The concentration paradox intensifies this policy argument. While scientific output is becoming more diffused worldwide, integrity issues are becoming concentrated in fewer countries and publishers. This opening gap indicates that monitoring of high-risk entities is extremely important. Variations in detection capabilities and reporting practices may play a role, but the Gini trajectories indicate that the concentration is real and increasing. 

Our findings should be interpreted with several important limitations in mind. We cannot definitively separate observed trends into changes in underlying misconduct versus changes in detection capacity. Periods of improved detection may artificially inflate retraction rates independently of actual integrity issues, while under-detection in certain publishers or countries (due to weaker screening, limited capacity, or language barriers) may artificially lower retraction rates. The decreasing publication-to-retraction lag (Appendix Fig.~\ref{fig:time to retraction}A: median dropping from $\sim$ 21 years in 1992 to $\sim$ 2 years recently) indicates faster detection, implying that at least part of the growth reflects better identification rather than rising misconduct. Even when limiting analysis to papers published before 2021, the lag distribution remains highly skewed with long tails. Papers from 2018-2021 may not be fully retracted yet. Visual inspection suggests potential acceleration in retractions during 2019-2021 across various categories; if this is due to short-term shocks rather than sustained exponential growth, our doubling time estimates might conflate different temporal patterns. Retraction flags may be missing, delayed, or incorrectly labeled. Coverage of conference proceedings is inconsistent, especially in computer science and engineering, which could bias field comparisons. Author disambiguation introduces variability, though overdispersed models help account for this. Retractions involve honest errors, misconduct, and policy changes. Practices around retractions differ across disciplines and publishers; we view the incidence as an operational indicator of stress on integrity rather than an exact measure of misconduct. Lastly, country-level patterns cannot distinguish between genuine misconduct, differential detection efforts, paper mill activity, or structural factors. Language bias is also possible: English-language journals may scrutinize non-native papers more intensely. 

Despite these limitations, our analysis offers the first principled, incidence-based statistical characterization of retraction dynamics. By modeling counts with exposure and validating exponential fits across various levels of aggregation, we show that per-paper and per-author risk have increased multiplicatively rather than simply following output growth. 
Three findings have direct policy implications. First, the system-wide multiplicative dynamics mean that small changes in the growth rate $g$ lead to large shifts in long-term risk. Interventions that modestly reduce $g$ can significantly alter outcomes over several years. Second, the concentration paradox shows that one-size-fits-all policies are not appropriate. Focused monitoring of high-risk countries, publishers, and fields is crucial. Third, some countries and publishers display flat or declining recent trends, showing that corrective measures are possible. 
The incidence framework presented here enables more accurate monitoring of the extent and growth rate over time. These metrics can evaluate whether reforms genuinely slow growth and reduce risk.

\section*{Methods}

\bmhead{Data sources \& coverage}
We analyze works from the October $2025$ snapshot of the open-source bibliometric dataset OpenAlex~\cite{priem2022openalex}.
Each work is tagged with one or more topics (4,516 total), organized into 252 subfields, 26 fields, and 4 domains. To avoid overcounting multi-topic papers, each work carries a unit weight that is split equally across its topics. If paper $p$ has $m_p$ topics, each topic receives $1/m_p$; the topic shares for $p$ sum to $1$. We then aggregate upward: a topic’s share contributes to its subfield, which in turn contributes to its field and domain. For any chosen cohort (e.g., a field or domain) and year $t$, the publication total $N_t$ and the retraction total $R_t$ are recomputed from these weights: $N_t$ is the sum of topic-based weights of all works published in year $t$ whose topics map into the cohort, and $R_t$ is the sum of the same weights restricted to retracted works. Hence, a work spanning multiple fields (rep. domains) contributes fractionally to each of the field (rep. domain).

We use an analogous fractional scheme for countries. Each work contributes a unit weight that is first split equally across its authors; an author’s share is then split equally across their affiliations. Affiliations are mapped to countries; country-level $N_t$ and $R_t$ in year $t$ are the sums of affiliation shares (for all works and for retracted works, respectively). This guarantees that a paper’s total contribution across countries sums to $1$. We report country results for the $21$ most productive countries in $2021$, which together account for more than $80\%$ of that year publications (see Fig.~\ref{fig:jitter}B).

Although each work lists a specific journal, we analyze the publisher level in order to have a better statistical signal. For each publisher and year $t$, we compute  $N_t$ and $R_t$ by summing all works (and retracted works) that the publisher issued in year $t$ (journal-to-publisher mapping provided by OpenAlex metadata). Incidence is then $i_t = R_t / N_t$ based on these  cohort totals. We focus on the top $20$ publishers by publication volume in $2021$, which together accounted for approximately $60\%$ of all publications that year. 
Note that some publishers were founded after $1992$. Specifically: Multidisciplinary Digital Publishing Institute ($1996$), Hindawi Publishing Corporation ($1997$), Medknow ($1997$), Lippincott Williams \& Wilkins ($1998$), BioMed Central ($2000$), and Frontiers Media ($2007$).

All fractional allocations are normalized to sum to $1$ per paper within each attribution layer (topics/fields/domains; authors/affiliations/countries), preventing double counting while preserving proportional credit. The totals $N_t$ and $R_t$ are always recomputed from the relevant weights for the cohort in question (field, domain, country, or publisher), so every incidence estimate reflects the appropriate denominator. Where metadata are missing (e.g., no country on an affiliation), the affected fraction is excluded from that aggregation.

\bmhead{Outcome, exposure, and incidence}
For each publication year $t$, we define an \emph{exposure} $N_t$ and an \emph{outcome} $R_t$, with both quantities instantiated to match the cohort under study. In the paper-level analyses, $N_t$ is the \emph{total number of works} published in year $t$ for a chosen cohort: e.g., the \emph{entire dataset}, a \emph{specific geography} (country/region), a \emph{scientific domain/field}, or a \emph{venue class} (publisher, journal family, mega-journal). The variable $R_t$ is the corresponding \emph{number of those works that are  retracted}. In author-level analyses, we analogously set $N_t \equiv A_t$ to be the \emph{number of active authors} in that cohort/year, i.e. those who published at least one work during that year, and $R_t$ to be the \emph{number of authors} with $\geq 1$ retraction attributable to that cohort/year. In all cases, the incidence (per-unit risk) is
$
i_t \;\equiv\; \mathbb{E}[R_t]/N_t.$
We analyze calendar-year cohorts from $1992$ to $2021$; the $2021$ cutoff limits right-censoring given publication-to-retraction lags (median $\approx 2.0$ years in the last decade; see Appendix).

\bmhead{Modeling Incidence with Exposure-Adjusted GLMs}
We model annual retraction counts using generalized linear models (GLMs)~\cite{nelder1972generalized} with an offset for exposure so that temporal effects pertain to \emph{incidence} rather than raw counts.\\ Formally,
$
R_t \mid N_t\sim \text{negative binomial, NB}(\lambda_t)$, and we have
$
\log \lambda_t = \eta_t +\log N_t,
$ where $\lambda_t \equiv \mathbb{E}[R_t]$ and $\eta_t$ parameterizes the evolution of incidence. 
We consider three different inference models as encapsulated in the $\eta_t$ function. 

\paragraph{(i) Constant incidence.}
\begin{equation}
i_t = i_0
\quad\Longleftrightarrow\quad
\eta_t = \alpha,
\end{equation}
so that $\log \lambda_t = \alpha + \log N_t$ and $i_0 = e^{\alpha}$ is the baseline incidence at $t_0$.

\paragraph{(ii) Linearly changing incidence.}
Here the rate itself drifts linearly:
\begin{equation}
i_t = i_0 + \beta\,(t-t_0)
\quad\Longleftrightarrow\quad
\lambda_t = N_t\big(i_0 + \beta\,(t-t_0)\big),
\end{equation}
fitted as a NB GLM with an identity link for the mean (equivalently, regress $R_t$ on $N_t$ and $N_t (t-t_0)$), enforcing $i_t>0$ over the estimation window.

\paragraph{(iii) Exponential-growth incidence.}
\begin{equation}
i_t = i_0\, e^{g\,(t-t_0)}
\quad\Longleftrightarrow\quad
\eta_t = \alpha + g\,(t-t_0),\;\; \alpha=\log i_0,
\end{equation}
so that $\log \lambda_t = \alpha + g\,(t-t_0) + \log N_t$. The parameter $g$ is the per-year growth rate and the doubling time is $T_d = \ln 2 / g$.\\

All three specifications are fit analogously at the author level by replacing $N_t$ with $A_t$ and the outcome with the annual count of authors with at least one retraction.
Analyses were conducted in \texttt{R} using standard NB-GLM implementations that estimate the dispersion parameter.

\bmhead{Model selection}
We compare the models using the Bayesian Information Criterion~\cite{schwarz1978estimating} to identify the best-fitting one, discarding models that either do not converge or yield parameters that are not statistically significant ($p\geq0.05$).
The Bayesian Information Criterion is a criterion for model selection among a finite set of candidate models.
The Bayesian Information Criterion of model $j$ is defined as:
$
    BIC_j = k  \ln n - 2 \ln \hat L_j,
$
where 
$n$ is the number of observations, which in our case corresponds to the $30$ years from $1992$ to $2021$;
$k$ is the number of parameters in the model,
and $\hat L_j$ is the maximized value of the likelihood function.
Lower BIC values indicate a better trade-off between fit and complexity; however, raw BIC values lack an immediate probabilistic interpretation. For model comparison, we therefore use \emph{BIC weights}. 
Given $\Delta_j= \mathrm{BIC}_j -\min_{\ell} \mathrm{BIC}_{\ell},
$ we then compute
\[
w_j \;=\; \frac{\exp\!\left(-\tfrac{1}{2}\,\Delta_j\right)}{\sum_{\ell} \exp\!\left(-\tfrac{1}{2}\,\Delta_{\ell}\right)} \, .
\]
The weights $w_j$ are non-negative and sum to $1$; under equal model priors and large-sample conditions, they can be interpreted as the models’ relative (approximate) posterior probabilities, i.e., the \emph{relative probability} that model $j$ is the best among the candidates, given the data. In the Appendix we provide the summary table of the model comparison for all the analysis reported.

\bmhead{Data availability}\label{Data Availability}
All data supporting the findings of this study are available. The primary bibliometric records were obtained from the OpenAlex October $2025$ snapshot. Retraction status flags used here are those distributed within OpenAlex and derived from the Retraction Watch database. Raw  records are not redistributed; they can be re-downloaded from OpenAlex.

\bmhead{Code availability}
The code used in the current study is available at \url{https://github.com/saraventurini/Retraction-Epidemic-in-Science}. The workflow uses Python (matplotlib, numpy, pandas, pycountry, rpy2, scipy) and R (glmmTMB, stats). Instructions to reproduce all results and “Source Data” are included in the repository README.


\printbibliography


\newpage
\appendix
\renewcommand{\thefigure}{A\arabic{figure}}
\setcounter{figure}{0}  
\renewcommand{\thetable}{A\arabic{table}}
\setcounter{table}{0}  
\appendix
\section*{Appendix}

\section{Global patterns of scientific production}
\begin{figure}[H]
\centering
\includegraphics[width=\textwidth]{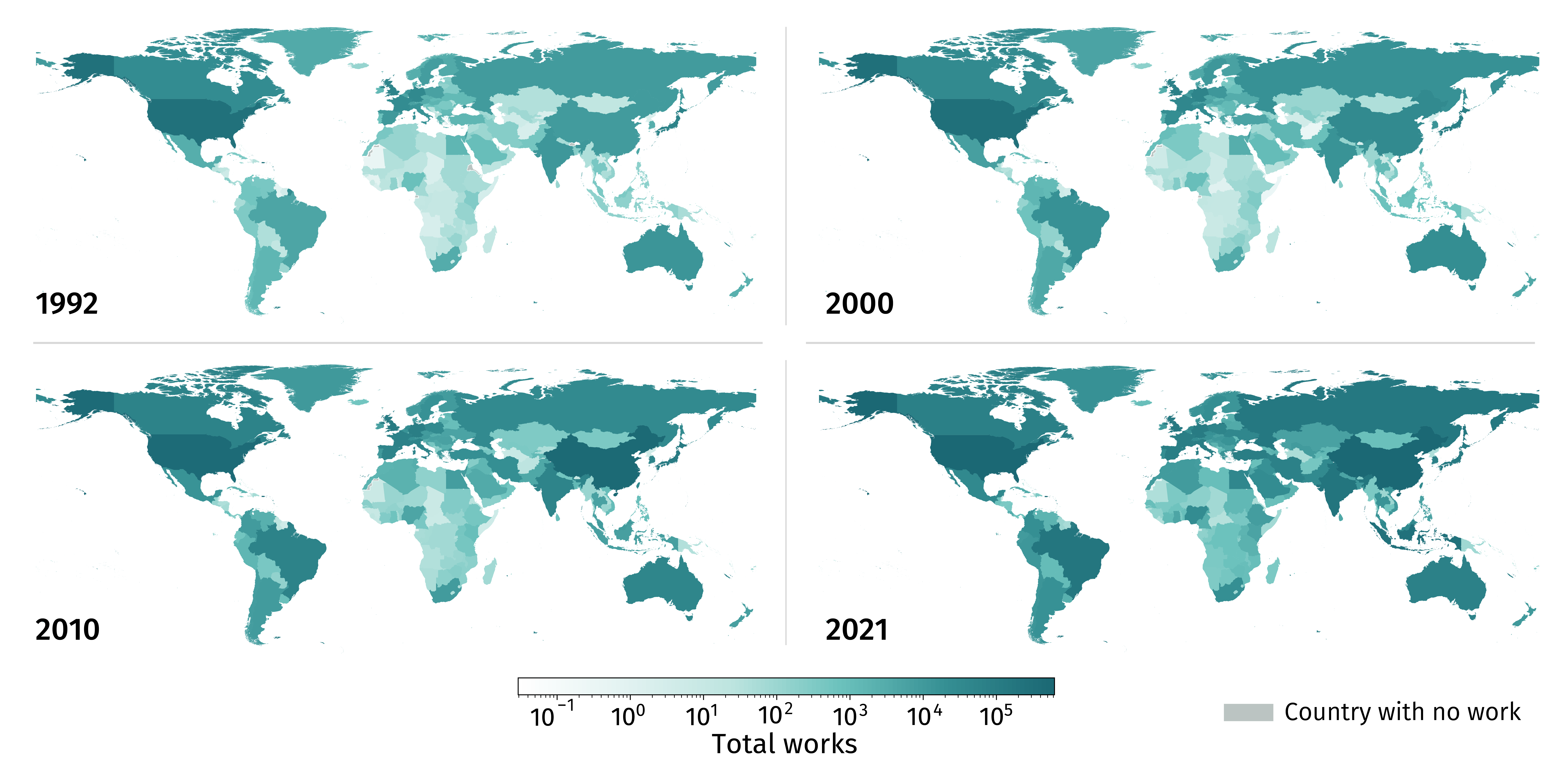}
\caption{
\textbf{Weighted number of published works per country.} 
Four world maps showing the geographic distribution of weighted publication counts in 1992, 2000, 2010, and 2021. Color intensity represents the number of works on a logarithmic scale (in gray countries with no publications).
}
\label{fig:global_works_sm}
\end{figure}

Maps in Fig.~\ref{fig:global_works_sm} illustrate the temporal growth and geographic expansion of published works over three decades. In early years ($1992$), scientific output was concentrated primarily in North America and Western Europe, with limited participation from Africa, South America, and parts of Asia. By $2000$ and $2010$, contributions increased and broadened geographically, reflecting expanding international research capacity. In $2021$, most countries exhibit measurable output, and several regions, including South America, East Asia, and Northern Africa, show substantial growth. Darker shading corresponds to higher total publication counts, while gray represents no recorded output. Overall, the figure highlights a marked global expansion of scientific productivity over time.

\section{Temporal Patterns in Retraction Dynamics}

We selected the pool of retracted works for analysis using the boolean retraction entry in OpenAlex. To examine temporal patterns in retraction dynamics, we matched more than $93\%$ of these works to the Retraction Watch dataset~\cite{retractionwatch2023}, which provides additional metadata including retraction dates.

In the main analysis, for each publication year we considered the pool of published works and determined how many were subsequently retracted in following years. Here we present two complementary analyses that provide additional insights into retraction timing.

\begin{figure}[H]
\centering
\includegraphics[width=\textwidth]{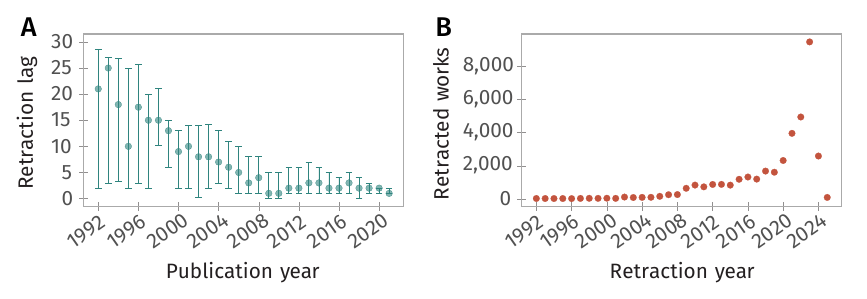}
\caption{
\textbf{Annual publication-to-retraction lags and retraction counts.} \textbf{A.} Median annual publication-to-retraction lags (circles); bars show the $25$th and $75$th percentiles. \textbf{B.} Annual counts of retracted works.}
\label{fig:time to retraction}
\end{figure}

First, for each publication year, we calculated the time elapsed between publication and retraction (publication-to-retraction lag). Fig.~\ref{fig:time to retraction}A shows that the median publication-to-retraction lag, calculated as the difference between retraction year and publication year, decreased markedly over time, from a median of approximately $21$ years in $1992$ to around $2$ years in the last decade. This indicates that more recent publications tend to be retracted more quickly after publication. This accelerating detection informed our decision to use $2021$ as the endpoint for our main analysis window, as works published after this date could had insufficient time for retraction processes to fully mature.

Second, for each year we counted the number of works for which retraction decisions were made in that year, regardless of their original publication date. This metric is useful for examining temporal changes in retraction policies and editorial practices. Fig.~\ref{fig:time to retraction}B shows that annual retraction counts increased consistently over time, reaching a peak in $2023$. The lower values observed for $2024$ and $2025$ may reflect incomplete data capture, as retraction notices for recent years may not yet be fully processed and indexed in the databases.

\section{Modeling Incidence with Exposure-Adjusted GLMs}

\subsection{Framework and Definitions}

In the main text, we consider a time window spanning $30$ years (from $1992$ to $2021$) and analyze the annual retraction incidence. For each publication year $t$, we record the total number of works published in year $t$, denoted $N_t$, and the number of those works that were subsequently retracted, denoted $R_t$. These quantities are computed for a chosen cohort: for example, the entire dataset, a specific geographic region, a scientific domain/field, or a venue class such as publisher. The observed annual retraction incidence (per-unit risk) is then defined as $R_t/N_t$, representing the proportion of works published in year $t$ that were eventually retracted.

In author-level analyses, we analogously record $A_t$ as the number of active authors in year $t$ (i.e., authors who published at least one work during that year), and count the number of those authors who have $\geq 1$ retracted work attributable to that year. The author-level incidence is then the ratio of these two quantities.

To model the temporal evolution of retraction incidence, we adopt a generalized linear model (GLM) framework standard in epidemiology. In this framework, $N_t$ (or $A_t$ for author-level analyses) represents the \emph{exposure}—the population at risk in year $t$—and $R_t$ represents the \emph{outcome}—the number of retraction events in that population. The incidence rate is then defined as $i_t = \mathbb{E}[R_t]/N_t$, where $\mathbb{E}[R_t]$ denotes the expected number of retractions under the model. The goal is to model how the incidence rate $i_t$ changes over time while accounting for variations in exposure.

\subsection{Generalized Linear Model Specification}
\label{sec:glm_modeling}

We model annual retraction counts $R_t$ using generalized linear models with an offset for exposure. This is a standard approach in epidemiology for modeling disease incidence in populations of varying size.

We assume that the observed retraction counts are generated from a counting process. Formally, we model $R_t$ using a negative binomial distribution, to account for overdispersion, where the dispersion parameter is estimated from the data. We denote $\lambda_t \equiv \mathbb{E}[R_t]$ as the expected number of retracted works in year $t$.

The GLM relates the expected count to the exposure and incidence through a log-link function:
\begin{equation}
\log \lambda_t = \eta_t + \log N_t,
\end{equation}
where $\eta_t$ parameterizes the evolution of incidence over time, and $\log N_t$ is the \emph{offset} term. The offset accounts for the varying population size at risk in each year, enforcing that changes in $N_t$ directly scale the expected count proportionally. The GLM estimates are obtained by maximizing the log-likelihood function. 

\subsection{Incidence Models}

We consider three different functional forms for $\eta_t$ to capture different hypotheses about how retraction incidence evolves over time:\\

\textbf{(i) Constant incidence.} The simplest model assumes that incidence remains stable over time:
\begin{equation}
i_t = i_0 \quad \Longleftrightarrow \quad \eta_t = \alpha,
\end{equation}
where $i_0 = e^\alpha$ is the baseline incidence. This implies $\log \lambda_t = \alpha + \log N_t$, so the expected count scales proportionally with exposure, but the per-capita rate remains constant.\\

\textbf{(ii) Linearly changing incidence.} Here the incidence rate drifts linearly with time:
\begin{equation}
i_t = i_0 + \beta (t - t_0) \quad \Longleftrightarrow \quad \lambda_t = N_t(i_0 + \beta(t - t_0)).
\end{equation}
Unlike the constant and exponential models (which use a log-link), the linear incidence model employs an identity link function. This means the expected count is modeled directly as a linear function of the predictors, rather than modeling its logarithm. Practically, this is fitted as a negative binomial GLM where $R_t$ is regressed on two predictors: $N_t$ (with coefficient $i_0$) and the interaction term $N_t(t - t_0)$ (with coefficient $\beta$). We constrain the fit to ensure $i_t > 0$ throughout the estimation window, as negative incidence rates are not meaningful.\\

\textbf{(iii) Exponential-growth incidence.} This model allows incidence to grow (or decline) exponentially:
\begin{equation}
i_t = i_0 e^{g(t - t_0)} \quad \Longleftrightarrow \quad \eta_t = \alpha + g(t - t_0), \quad \alpha = \log i_0,
\end{equation}
where $\log \lambda_t = \alpha + g(t - t_0) + \log N_t$. The parameter $g$ is the per-year growth rate.
For exponential growth ($g > 0$), we can characterize the speed of increase using the \emph{doubling time} $T_d = \ln 2/g$, defined as the time required for the incidence to double.\\

All three specifications are fit analogously at the author level by replacing $N_t$ with $A_t$ and the outcome with the annual count of authors with at least one retraction. 

Analyses were conducted in R using standard GLMM implementations from the \texttt{glmmTMB} package.

\subsection{Model Selection}

To compare the three incidence models, we use the Bayesian Information Criterion (BIC), which balances model fit with complexity. For each model $j$, we calculate:
\begin{equation}
BIC_j = k  \ln n - 2 \ln \hat L_j,
\end{equation}
where $\hat L_j$ is the maximized log-likelihood, $k$ is the number of parameters, and $n$ is the number of observations.
In our analysis, $n = 30$ corresponds to the number of years in the time window ($1992\text{-}2021$). The number of parameters $k$ differs across models:
constant incidence has $k = 2$ (parameters $\alpha$, and dispersion parameter $\gamma$),
linearly changing incidence has $k = 3$ ($i_0$, $\beta$, and dispersion parameter $\gamma$),
and exponential-growth incidence: $k = 3$ (parameters $\alpha$, $g$, and dispersion parameter $\gamma$). 

We then compute BIC weights to quantify the relative support for each model:
\[
w_j \;=\; \frac{\exp\!\left(-\tfrac{1}{2}\,\Delta_j\right)}{\sum_{\ell} \exp\!\left(-\tfrac{1}{2}\,\Delta_{\ell}\right)} \, .
\]
where $\Delta_j = \text{BIC}_j - \min_{\ell}(\text{BIC}_{\ell})$. These weights can be interpreted as the approximate probability that each model is the best among the candidate set, given the data.

\subsection{Reporting of Results}

Throughout this Appendix, we present results in a standardized format to facilitate interpretation and comparison across different cohorts (entire dataset, domains, fields, countries, publishers).\\

\textbf{Model comparison tables.} For each cohort, we report BIC weights (in percentage) for the three incidence models: constant incidence (constant), linearly changing incidence (linear), and exponential-growth incidence (exponential). The highest BIC weights are highlighted in bold. Models that did not converge or for which the parameter $p$-values were not statistically significant ($P > 0.05$) are indicated with a dash. These tables allow quick identification of which model best describes the data for each cohort.\\

\textbf{Parameter estimate tables.} For cohorts where the exponential-growth model achieves a BIC weight higher than  $90\%$, we report
the optimized dispersion parameter $\gamma$ and the annual growth rate ($g$) with its associated $p$-value (using standard significance notation: * $P \leq 0.05$; ** $P \leq 0.01$; *** $P \leq 0.001$; **** $P \leq 0.0001$), and the doubling time ($T_d$, in years) with its 95\% confidence interval in brackets. The confidence intervals for $T_d$ are obtained by propagating the uncertainty in the estimated growth rate $g$ through the transformation $T_d = \ln 2 / g$.\\

\textbf{Figures.} We present two types of visualizations depending on model fit quality. 
For cohorts with strong exponential-growth fits (BIC weight $>90\%$), figures show the observed annual retraction incidence (data points), the fitted GLM exponential-growth curve (line), and $95\%$ confidence intervals (shaded areas) on a lin-log scale (logarithmic $y$-axis). An inset plot displays the same data on a lin-lin scale (linear $y$-axis) with 95\% prediction intervals. 
For cohorts without strong exponential fits, figures show the observed annual retraction incidence (data points) along with a $5$-year rolling average (line) and its corresponding standard deviation (shaded areas).

\section{Sensitivity Analyses}

In this section, we report two sensitivity analyses: one regarding the time window under analysis and another concerning the overdispersion assumption of the data.\\

\subsection{Sensitivity to Time Window Selection} To assess whether our findings depend on the specific choice of time window, we repeat the GLM analysis using alternative time periods: $1992{-}2017$, $1992{-}2019$, $1992{-}2022$, $1997{-}2021$, and $2002\text{-}2021$.

Table~\ref{tbl:works_ranges} presents BIC weights for the three incidence models across these different time windows. The exponential-growth model consistently achieves $100\%$ BIC weight for all time periods examined, demonstrating robust support regardless of the window chosen.

Table~\ref{tbl:works} reports the corresponding dispersion parameters, growth rates and doubling times for each time window. Growth rates range from $0.11$ to $0.13$ per year with doubling times between $5.4$ and $6.1$ years, confirming the stability of our estimates across different time windows. 

The consistency of these results indicates that the observed exponential growth pattern is not an artifact of the specific time period analyzed.

{\footnotesize
\renewcommand{\arraystretch}{0.75}
\setlength{\tabcolsep}{3pt}
\begin{table}[H]
    \centering
    \begin{tabular}{l|ccc}
\textbf{Range/Fit} & \textbf{Constant} & \textbf{Linear} & \textbf{Exponential} \\ \midrule
1992-2017 & 0.0\% & 0.0\% & \textbf{100.0\%} \\
1992-2019 & 0.0\% & 0.0\% & \textbf{100.0\%} \\
1992-2021 & 0.0\% & 0.0\% & \textbf{100.0\%} \\
1992-2022 & 0.0\% & 0.0\% & \textbf{100.0\%} \\
1997-2021 & 0.0\% & 0.0\% & \textbf{100.0\%} \\
2002-2021 & 0.0\% & 0.0\% & \textbf{100.0\%} \\
    \end{tabular}
    \caption{\textbf{GLM fits of retraction incidence under different growth models.} The table shows the BIC weights in percentage, with the highest weights highlighted in bold. Rows correspond to different time windows (in the main text results for the $30$ years window from $1992$ to $2021$ are shown), and columns correspond to the incidence growth models: constant incidence (constant), linearly changing incidence (linear), and exponential-growth incidence (exponential). Models that did not converge or for which the parameter p-values were not statistically significant ($P \geq 0.05$) are indicated with a dash.}
    \label{tbl:works_ranges}
\end{table}
}

{\footnotesize
\renewcommand{\arraystretch}{0.75}
\setlength{\tabcolsep}{3pt}
\begin{table}[H]
    \centering
    \begin{tabular}{l|lll}
\textbf{Range} & $\boldsymbol{\alpha}$ & $\mathbf{g}$ & \textbf{$\mathbf{T_d}$ (years)}  \\ \midrule
1992-2017 & 0.03 & 0.12 (****) & 5.8 [5.3-6.4] \\
1992-2019 & 0.03 & 0.12 (****) & 5.9 [5.4-6.5] \\
1992-2021 & 0.04 & 0.12 (****) & 5.6 [5.1-6.0] \\
1992-2022 & 0.05 & 0.13 (****) & 5.4 [5.0-5.8] \\
1997-2021 & 0.04 & 0.12 (****) & 5.7 [5.2-6.4] \\
2002-2021 & 0.05 & 0.11 (****) & 6.1 [5.3-7.2] \\
\end{tabular}
    \caption{\textbf{GLM fits of publication retraction incidence using an exponential-growth model.}
    Each row corresponds to a different time window. Reported are the estimated dispersion parameter ($\gamma$), the annual growth rate ($g$) with the associated $p$-value, and the doubling time ($T_d$, in years) with its 95\% CI in brackets.
    Significance levels: *~$P~\leq~0.05$; **~$P~\leq~0.01$; ***~$P~\leq~0.001$; ****~$P~\leq~0.0001$.}
    \label{tbl:works}
\end{table}
}

\subsection{Robustness to Overdispersion Assumptions} 
In the main text, we use GLMs with a negative binomial (NB) likelihood rather than a Poisson likelihood to account for overdispersion in the retraction count data. The choice between these two distributions depends on the variance structure of the data. For a Poisson distribution, the variance equals the mean: $\sigma^2 = \mu$. However, count data often exhibit overdispersion, where the variance exceeds the mean. The negative binomial distribution accommodates this through an additional dispersion parameter $\gamma$, with variance $\sigma^2 = \mu + \gamma \mu^2$. As $\gamma \to 0$, the negative binomial converges to the Poisson distribution; higher values of $\gamma$ indicate stronger overdispersion.

Here, we provide empirical evidence supporting the use of the negative binomial model through a mean-variance analysis and examine the sensitivity of our results to the dispersion parameter.

\begin{figure}[H]
\centering
\includegraphics[width=\textwidth]{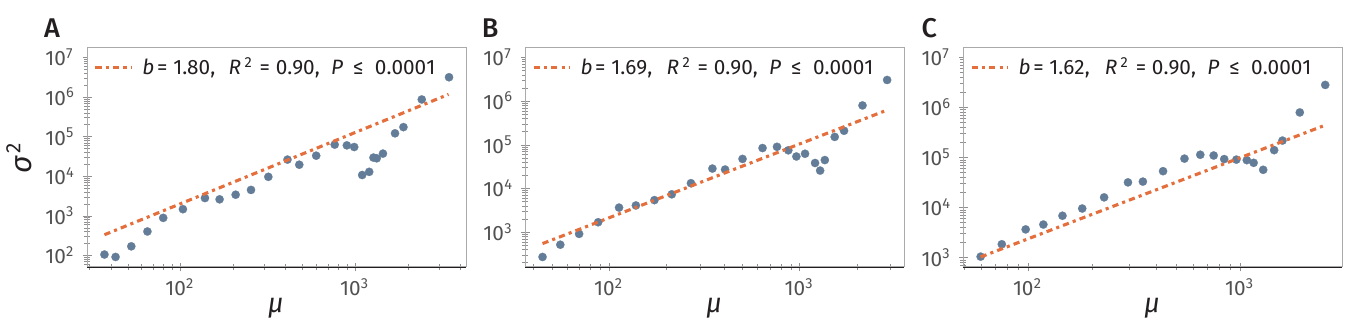}
\caption{\textbf{Overdispersion analysis.}
Scaling behavior between the mean (\(\mu\)) and variance (\(\sigma^2\)) of retraction counts across sliding windows. 
Blue circles represent the empirical mean–variance data, and the orange dotted line indicates the fitted power-law relationship \(\sigma^2 \propto \mu^b\). 
The legend reports the estimated slope (\(b\)), coefficient of determination (\(R^2\)), and corresponding \(P\)-value. 
Panels correspond to different window sizes: $5$ years (\textbf{A}); $7$ years (\textbf{B}); $9$ years (\textbf{C}).
}
\label{fig:overdispersion}
\end{figure}

We filtered the dataset to include only the years from $1992$ to $2021$, then constructed sliding windows of fixed sizes $s \in \{5, 7, 9\}$ years. For each window, we extracted the yearly counts of retracted works and computed both the mean ($\mu$) and the variance ($\sigma^2$). Fig.~\ref{fig:overdispersion} shows the fitted scaling behavior, which reveals a power-law mean--variance relationship of the form $\sigma^2 \propto \mu^b$. The estimated slopes $b$ ranged from approximately $1.60$ to $1.80$ across window sizes, indicating non-linear scaling consistent with overdispersion and multiplicative dynamics. A slope of $b = 1$ would indicate Poisson-like behavior (variance equals mean), while $b > 1$ indicates overdispersion, supporting the use of the negative binomial model.
We estimate the dispersion parameter $\gamma$ from the data rather than fixed a priori, and report the estimates in the Appendix Tables.

\section{Exponential Growth of Authors Retraction Incidence}
\begin{figure}[H]
\centering
\includegraphics[width=\textwidth]{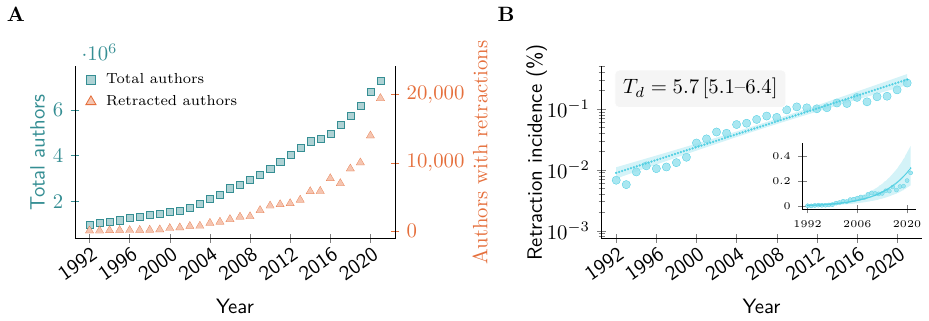}
\caption{
\textbf{Temporal patterns of authors retractions.}
\textbf{A.} Annual number of active authors (squares, green) and active authors with $>1$ retractions (triangles, orange).  
\textbf{C.} Annual authors retraction incidence in percentage (circles, blue), with the GLM exponential-growth incidence fit based on a negative binomial likelihood (blue line, growth rate $P \leq 0.0001$), and its $95\%$ confidence interval (shaded area). The doubling time ($T_d$) with its $95\%$ CI is reported in the legend. Values are shown on a logarithmic $y$-axis (linear scale with 95\% prediction intervals in inset).}
\label{fig:authors}
\end{figure}

In addition to analyzing the temporal dynamics of retractions at the publication level, we also considered them at the author level.
For each publication year $t$, we recorded the total number of active authors in that year (i.e., authors who published at least one work) and the subset of authors with $\geq1$ retracted works.
In Fig.~\ref{fig:authors}A, we plot, for each year $t$ from $1992$ to $2021$, the total number of active authors and the number with at least one retraction. As in the case of publications, both series increase over time.
In Fig.~\ref{fig:authors}B, we show that the share of active authors with $\geq1$ retraction in a given year also grows exponentially in the last $30$ years. 
We estimated incidence trends using the same exposure-adjusted GLM framework applied at the publication level. In this case as well, the evidence strongly favors the exponential-growth incidence model.
Using the estimated growth rate from a GLM with exponential growth and a negative binomial likelihood,  we obtain a doubling time of $5.6$ years ($95\%$ CI $[4.2\text{–}8.7]$), similar to the paper-level rate.
We conduct robustness analysis across different time periods (Table~\ref{tbl:authors_ranges}) and report the corresponding growth rates and doubling times (Table~\ref{tbl:authors}).
The exponential-growth model achieves 100\% BIC weight across all tested time windows, with growth rates consistently ranging from $0.08$ to $0.13$ per year and doubling times between $5.2$ and $8.3$ years.
This stability confirms that our findings are robust to the choice of time period.

{\footnotesize
\renewcommand{\arraystretch}{0.75}
\setlength{\tabcolsep}{3pt}
\begin{table}[H]
    \centering
    \begin{tabular}{l|ccc}
\textbf{Range/Fit} & \textbf{Constant} & \textbf{Linear} & \textbf{Exponential} \\ \midrule
1992-2017 & 0.0\% & 0.0\% & \textbf{100.0\%} \\
1992-2019 & 0.0\% & 0.1\% & \textbf{99.9\%} \\
1992-2021 & 0.0\% & 0.0\% & \textbf{100.0\%} \\
1992-2022 & 0.0\% & 0.0\% & \textbf{100.0\%} \\
1997-2021 & 0.0\% & - & \textbf{100.0\%} \\
2002-2021 & 0.0\% & 0.0\% & \textbf{100.0\%} \\
    \end{tabular}
    \caption{\textbf{GLM fits of retraction incidence at author-level under different growth models.} The table shows the BIC weights in percentage, with the highest weight highlighted in bold. Rows correspond to different time windows, and columns correspond to the incidence growth models: constant incidence (constant), linearly changing incidence (linear), and exponential-growth incidence (exponential). Models that did not converge or for which the parameter p-values were not statistically significant ($P > 0.05$) are indicated with a dash.}
    \label{tbl:authors_ranges}
\end{table}
}

{\footnotesize
\renewcommand{\arraystretch}{0.75}
\setlength{\tabcolsep}{3pt}
\begin{table}[H]
    \centering
    \begin{tabular}{l|lll}
\textbf{Range} & $\boldsymbol{\alpha}$ & $\mathbf{g}$ & \textbf{$\mathbf{T_d}$ (years)}  \\ \midrule
1992-2017 & 0.06 & 0.13 (****) & 5.2 [4.7-5.9] \\
1992-2019 & 0.07 & 0.13 (****) & 5.5 [4.9-6.2] \\
1992-2021 & 0.07 & 0.12 (****) & 5.7 [5.1-6.4] \\
1992-2022 & 0.07 & 0.12 (****) & 5.7 [5.2-6.3] \\
1997-2021 & 0.05 & 0.11 (****) & 6.5 [5.7-7.5] \\
2002-2021 & 0.01 & 0.08 (****) & 8.3 [7.5-9.4] \\
\end{tabular}
    \caption{\textbf{GLM fits of retraction incidence at author-level using an exponential-growth model.}
    estimated dispersion parameter
    Each row corresponds to a different time window. Reported are the estimated dispersion parameter ($\gamma$), the annual growth rate ($g$) with the associated $p$-value, and the doubling time ($T_d$, in years) with its 95\% CI in brackets.
    Significance levels: *~$P~\leq~0.05$; **~$P~\leq~0.01$; ***~$P~\leq~0.001$; ****~$P~\leq~0.0001$.}
    \label{tbl:authors}
\end{table}
}

\section{Retraction Incidence by Research Domain and Field}

We extend our analysis to examine retraction incidence at the domain and field level using topic tags from OpenAlex. Each work in OpenAlex is tagged with one or more topics; we exclude works without topic information from this analysis.


To avoid double-counting when works are assigned to multiple topics, we apply fractional weighting. Each work $p$ carries a unit weight that is split equally across its $m_p$ topics: each topic receives a weight of $1/m_p$, ensuring that the topic shares for work $p$ sum to 1. These weights aggregate hierarchically: a topic's share contributes to its parent subfield, which contributes to its parent field, which contributes to its parent domain. 

For any chosen cohort (e.g., a specific field or domain) and year $t$, we recompute the publication total $N_t$ and the retraction total $R_t$ from these fractional weights. Specifically, $N_t$ is the sum of topic-based weights of all works published in year $t$ whose topics map into the cohort, and $R_t$ is the analogous sum restricted to retracted works. Consequently, a work spanning multiple fields (or domains) contributes fractionally to each. 

We then apply the same GLM framework to these fractionally-weighted cohorts, computing BIC weights, dispersion parameters, growth rates, and doubling times for each domain and field.

\begin{figure}[H]
\centering
\includegraphics[width=0.7\textwidth]{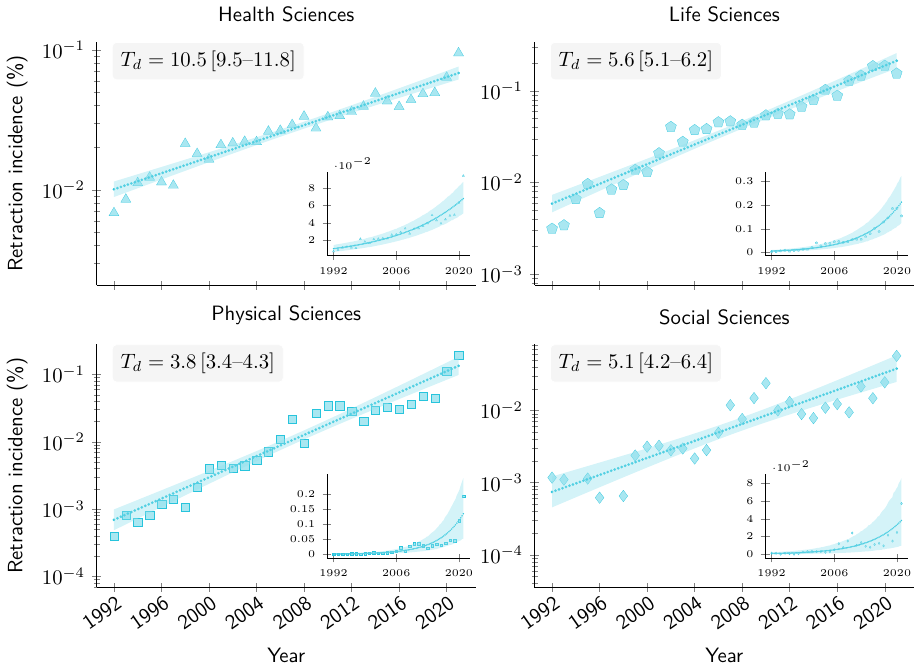} 
    \caption{\textbf{
    Retraction incidence over time across domains.} 
    Annual retraction incidence (blue) by domain (Health Sciences — triangles; Life Sciences — pentagons; Physical Sciences — squares; Social Sciences, diamonds), with the GLM exponential-growth incidence fit based on a negative binomial likelihood (blue line, 
    growth rate significance levels $P \leq 0.05$ 
    ), and its $95\%$ confidence interval (shaded area). The doubling time ($T_d$) with its $95\%$ CI is reported in the legend. Values are shown on a logarithmic $y$-axis (linear scale with 95\% prediction intervals in inset).}
    \label{fig:domains2}
\end{figure}

In Fig.~\ref{fig:domains2} (and in Fig.~\ref{fig:domains}A in the main text), we show the GLM exponential-growth incidence fits based on a negative binomial likelihood for the four domains. Table~\ref{tbl:domains_} shows that, according to the BIC weights, the exponential-growth incidence model provides the best fit among the three models considered, achieving $100\%$ BIC weight across all four domains.
Table~\ref{tbl:domains} reports the estimated dispersioon parameters and incidence growth rates, and corresponding doubling times, which vary substantially across domains. Health Sciences exhibits the slowest growth with a doubling time of $10.5$ years ($95\%$ CI $[9.5\text{-}11.8]$), while Physical Sciences shows the fastest growth with a doubling time of $3.8$ years ($95\%$ CI $[3.4\text{-}4.3]$). Life Sciences and Social Sciences have intermediate doubling times of $5.6$ years ($95\%$ CI $[5.1\text{-}6.2]$) and $5.1$ years ($95\%$ CI $[4.2\text{-}6.4]$), respectively.

To examine heterogeneity not only across the four domains but also within them, we extend our analysis to the second level of the OpenAlex topic hierarchy, which consists of $26$ fields nested within the domains.
Table~\ref{tbl:fields_} shows the comparison of the three incidence models using BIC weights across the different fields. We obtain consistent evidence supporting exponential growth across all $26$ fields. The exponential-growth model achieves a BIC weight higher than $90\%$ in $24$ out of $26$ fields. In Veterinary the constant-incidence model retains a non-negligible weight ($11.9\%$), although the exponential-growth model remains strongly favored ($88.1\%$). In Psychology, the linear model receives minimal support ($0.4\%$) while the exponential model dominates ($99.6\%$). For many fields, the linear-incidence model did not converge or its parameters were not statistically significant ($P > 0.05$).

Table~\ref{tbl:fields} reports the corresponding dispersion parameters, growth rates and doubling times. The results reveal marked heterogeneity across fields. Within Health Sciences, doubling times range from $4.9$ years in Health Professions to $10.8$ years in Medicine. Within the Life Sciences, growth rates vary substantially, with doubling times ranging from $5.0$ years (Biochemistry, Genetics and Molecular Biology) to $11.7$ years (Immunology and Microbiology). Physical Sciences exhibits the most rapid growth overall: Computer Science has the shortest doubling time at $2.8$ years, while most other fields in this domain have doubling times between $3.7$ and $6.0$ years. Social Sciences displays intermediate growth rates, with doubling times ranging from $4.2$ years (Decision Sciences) to $7.9$ years (Psychology).
These field doubling times are also displayed in Fig.~\ref{fig:domains}B in the main text; Table.~\ref{tbl:abbreviations} provides the corresponding abbreviations.

\begin{figure}[H]
\centering
\includegraphics[width=\textwidth]{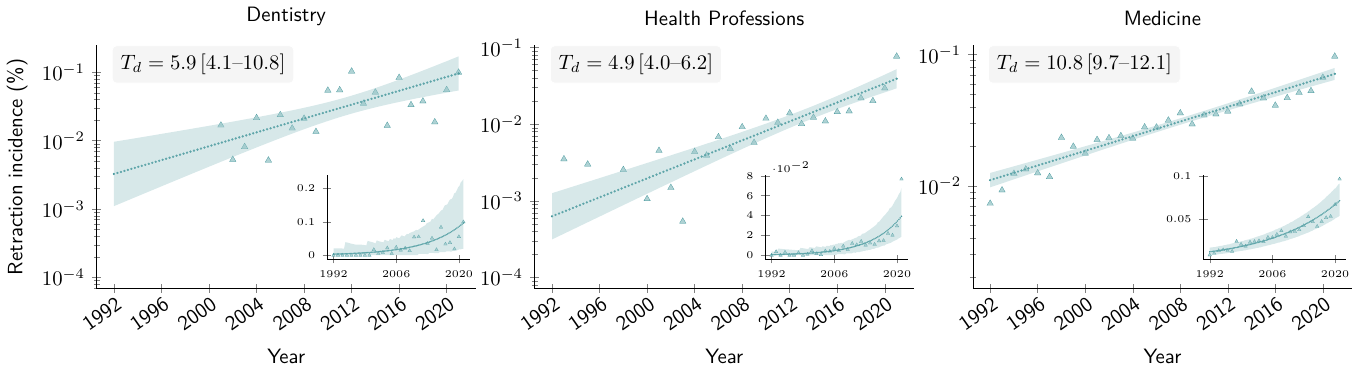} 
    \caption{\textbf{
    Retraction incidence over time across fields within the Health Sciences domain and with a strongly best-fit GLM exponential-growth.} Each panel representing one of the fields in Health Sciences domain with a strongly best-fit GLM exponential-growth. Green triangles show the annual retraction incidence (in percentage), and blue lines represent the GLM exponential-growth incidence fits based on a negative binomial likelihood (growth rate significance levels $P \leq 0.05$) with shaded areas indicating the $95\%$ confidence intervals. The doubling times ($T_d$) with their $95\%$ CI are reported in the legend. The $y$-axis is shown on a logarithmic scale (zero values not shown; linear scale with 95\% prediction intervals in inset).}
    \label{fig:health sciences}
\end{figure}

\begin{figure}[H]
\centering
\includegraphics[width=\textwidth]{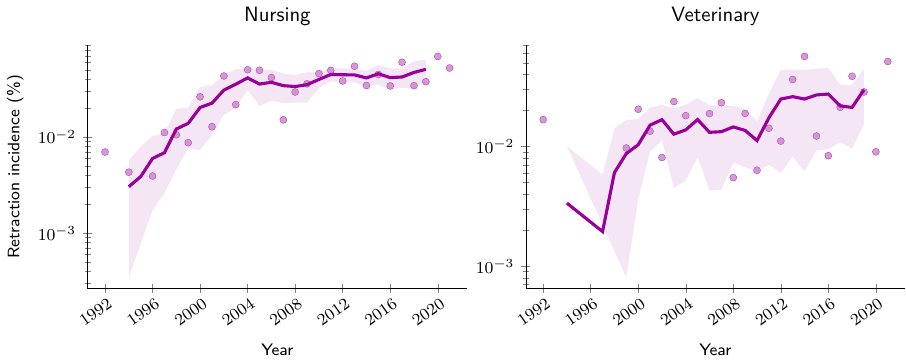} 
    \caption{\textbf{
    Retraction incidence over time across fields within the Health Sciences domain without a strongly best-fit GLM exponential-growth incidence model.} Each panel represents one of the fields within the Health Sciences domain for which no strong exponential-growth fit was observed. 
    Purple circles indicate the annual retraction incidence (in percentage), while blue lines show the centered $5$-year rolling average, with shaded areas representing the corresponding standard deviation.
    }
    \label{fig:health sciences no exponential}
\end{figure}

\begin{figure}[H]
\centering
\includegraphics[width=\textwidth]{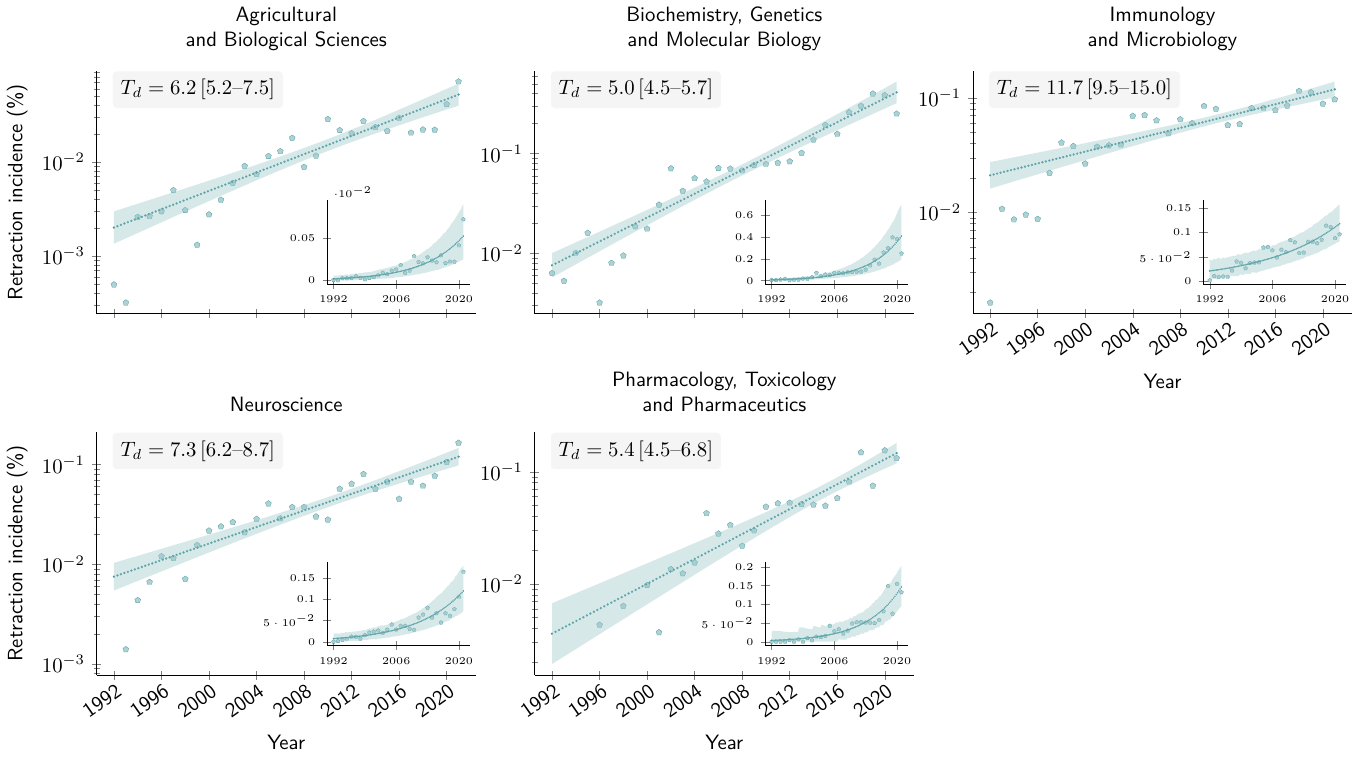}
    \caption{\textbf{
    Retraction incidence over time across fields within the Life Sciences domain.} Each panel representing one of the fields in Health Sciences domain. Green triangles show the annual retraction incidence (in percentage), and blue lines represent the GLM exponential-growth incidence fits based on a negative binomial likelihood (growth rate significance levels $P \leq 0.05$) with shaded areas indicating the $95\%$ confidence intervals. The doubling times ($T_d$) with their $95\%$ CI are reported in the legend. The $y$-axis is shown on a logarithmic scale (zero values not shown; linear scale with 95\% prediction intervals in inset).}
    \label{fig:life sciences}
\end{figure}

\begin{figure}[H]
\centering
\includegraphics[width=\textwidth]{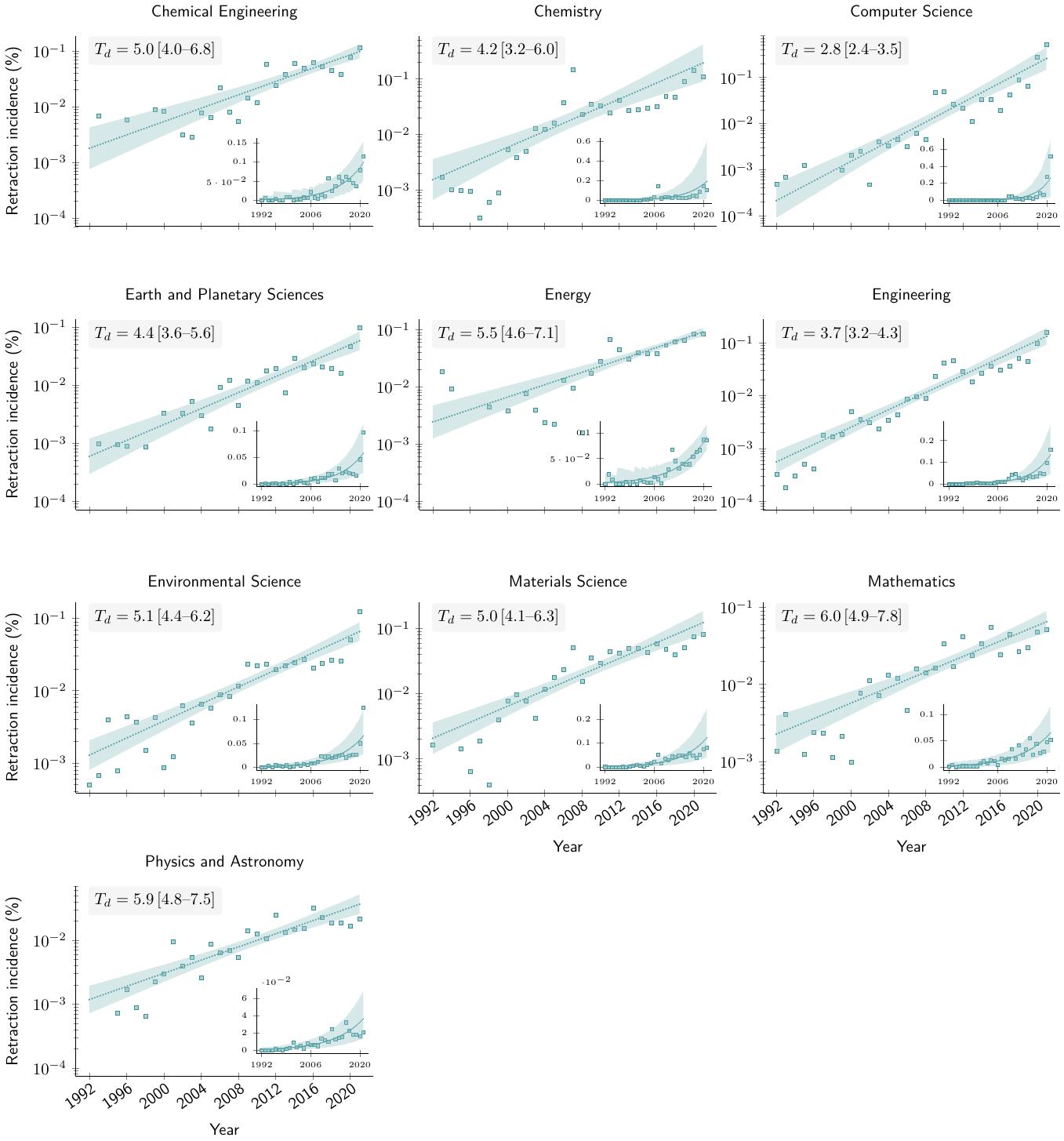}
    \caption{\textbf{
    Retraction incidence over time across fields within the Physical Sciences domain.} Each panel representing one of the fields in Health Sciences domain. Green triangles show the annual retraction incidence (in percentage), and blue lines represent the GLM exponential-growth incidence fits based on a negative binomial likelihood (growth rate significance levels $P \leq 0.05$),
    with shaded areas indicating the $95\%$ confidence intervals. The doubling times ($T_d$) with their $95\%$ CI are reported in the legend. The $y$-axis is shown on a logarithmic scale (zero values not shown; linear scale with 95\% prediction intervals in inset).}
    \label{fig:physical sciences}
\end{figure}

\begin{figure}[H]
\centering
\includegraphics[width=\textwidth]{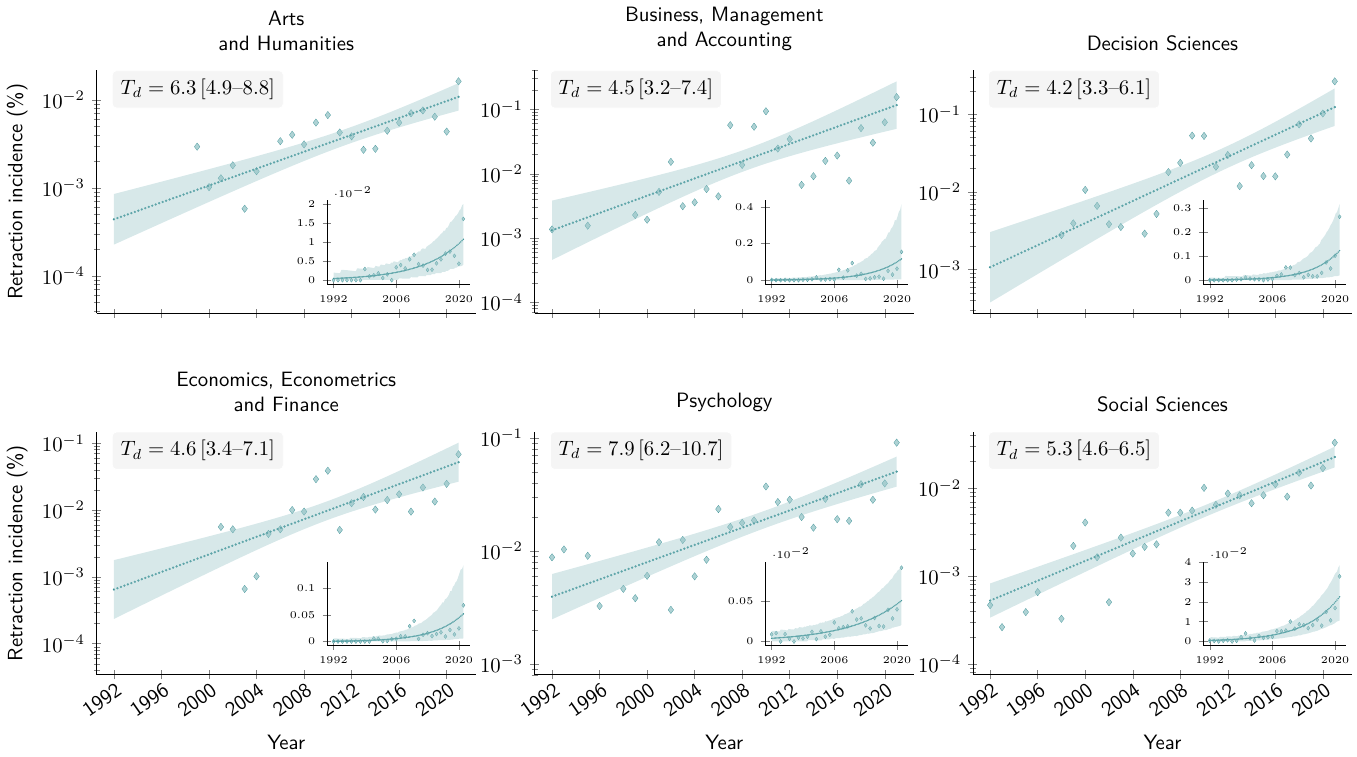}
    \caption{\textbf{
    Retraction incidence over time across fields within the Social Sciences domain.} Each panel representing one of the fields in Health Sciences domain. Blue triangles show the annual retraction incidence (in percentage), and blue lines represent the GLM exponential-growth incidence fits based on a negative binomial likelihood  (
    growth rate significance levels $P \leq 0.05$
    ) with shaded areas indicating the $95\%$ confidence intervals. The doubling times ($T_d$) with their $95\%$ CI are reported in the legend. The $y$-axis is shown on a logarithmic scale (zero values not shown; linear scale with 95\% prediction intervals in inset).}
    \label{fig:social sciences}
\end{figure}

Figures~\ref{fig:health sciences}--\ref{fig:social sciences} show the GLM exponential-growth incidence fits based on a negative binomial likelihood for each field, illustrating the consistent upward trends across all disciplines despite the variation in growth rates.
Fig.~\ref{fig:health sciences no exponential} displays retraction incidence patterns for the remaining fields of Nursing and Veterinary.

{\footnotesize
\renewcommand{\arraystretch}{0.75}
\setlength{\tabcolsep}{3pt}
\begin{table}[H]
    \centering
    \begin{tabular}{l|ccc}
\textbf{Domain / Fit} & \textbf{Constant} & \textbf{Linear} & \textbf{Exponential} \\ \midrule
Health Sciences & 0.0\% & 0.0\% & \textbf{100.0\%} \\
Life Sciences & 0.0\% & 0.0\% & \textbf{100.0\%} \\
Physical Sciences & 0.0\% & - & \textbf{100.0\%} \\
Social Sciences & 0.0\% & - & \textbf{100.0\%} \\
\end{tabular}
    \caption{\textbf{GLM fits of retraction incidence at domain-level under different growth models.} The table shows the BIC weights in percentage, with the highest weight highlighted in bold. Rows correspond to the $4$ OpenAlex domains, and columns correspond to the incidence growth models: constant incidence (constant), linearly changing incidence (linear), and exponential-growth incidence (exponential). Models that did not converge or for which the parameter p-values were not statistically significant ($P > 0.05$) are indicated with a dash.}
    \label{tbl:domains_}
\end{table}
}

{\footnotesize
\renewcommand{\arraystretch}{0.75}
\setlength{\tabcolsep}{3pt}
\begin{table}[H]
    \centering
    \begin{tabular}{l|lll}
\textbf{Domain} & $\boldsymbol{\gamma}$ & $\mathbf{g}$ & \textbf{$\mathbf{T_d}$ (years)}  \\ \midrule
Health Sciences & 0.02 & 0.07 (****) & 10.5 [9.5-11.8] \\
Life Sciences & 0.06 & 0.12 (****) & 5.6 [5.1-6.2] \\
Physical Sciences & 0.15 & 0.18 (****) & 3.8 [3.4-4.3] \\
Social Sciences & 0.28 & 0.14 (****) & 5.1 [4.2-6.4] \\
\end{tabular}
    \caption{\textbf{GLM fits of retraction incidence at domain-level using an exponential-growth model.}
    Each row corresponds to one of the four domains. Reported are the estimated dispersion parameter ($\gamma$), the annual growth rate ($g$) with the associated $p$-value, and the doubling time ($T_d$, in years) with its 95\% CI in brackets.
    Significance levels: *~$P~\leq~0.05$; **~$P~\leq~0.01$; ***~$P~\leq~0.001$; ****~$P~\leq~0.0001$.}
    \label{tbl:domains}
\end{table}
}

{\footnotesize
\renewcommand{\arraystretch}{0.75}
\setlength{\tabcolsep}{3pt}
\begin{table}[H]
    \centering
    \begin{tabular}{ll|ccc}
\textbf{Domain} & \textbf{Field/Fit} & \textbf{Constant} & \textbf{Linear} & \textbf{Exponential} \\ \midrule
\multirow{3}{*}{\makecell{Health\\Sciences}} & Dentistry & 0.0\% & - & \textbf{100.0\%} \\
  & Health Professions & 0.0\% & - & \textbf{100.0\%} \\
  & Medicine & 0.0\% & 0.0\% & \textbf{100.0\%} \\
  & Nursing & \textbf{100.0\%} & - & - \\
  & Veterinary & 11.9\% & - & \textbf{88.1\%} \\
 \midrule
\multirow{3}{*}{\makecell{Life\\Sciences}}  & Agricultural and Biological Sciences & 0.0\% & - & \textbf{100.0\%} \\
  & Biochemistry, Genetics and Molecular Biology & 0.0\% & 0.0\% & \textbf{100.0\%} \\
  & Immunology and Microbiology & 0.0\% & - & \textbf{100.0\%} \\
  & Neuroscience & 0.0\% & - & \textbf{100.0\%} \\
  & Pharmacology, Toxicology and Pharmaceutics & 0.0\% & - & \textbf{100.0\%} \\
 \midrule
\multirow{3}{*}{\makecell{Physical\\Sciences}}  & Chemical Engineering & 0.0\% & - & \textbf{100.0\%} \\
  & Chemistry & 0.0\% & - & \textbf{100.0\%} \\
  & Computer Science & 0.0\% & - & \textbf{100.0\%} \\
  & Earth and Planetary Sciences & 0.0\% & - & \textbf{100.0\%} \\
  & Energy & 0.0\% & - & \textbf{100.0\%} \\
  & Engineering & 0.0\% & - & \textbf{100.0\%} \\
  & Environmental Science & 0.0\% & - & \textbf{100.0\%} \\
  & Materials Science & 0.0\% & - & \textbf{100.0\%} \\
  & Mathematics & 0.0\% & - & \textbf{100.0\%} \\
  & Physics and Astronomy & 0.0\% & - & \textbf{100.0\%} \\
 \midrule
\multirow{3}{*}{\makecell{Social\\Sciences}} & Arts and Humanities & 0.0\% & - & \textbf{100.0\%} \\
  & Business, Management and Accounting & 0.0\% & - & \textbf{100.0\%} \\
  & Decision Sciences & 0.0\% & - & \textbf{100.0\%} \\
  & Economics, Econometrics and Finance & 0.0\% & - & \textbf{100.0\%} \\
  & Psychology & 0.0\% & 0.4\% & \textbf{99.6\%} \\
  & Social Sciences & 0.0\% & - & \textbf{100.0\%} \\
\end{tabular}
    \caption{\textbf{GLM fits of retraction incidence at field-level under different growth models.} The table shows the BIC weights in percentage, with the highest weight highlighted in bold. Rows correspond to the $26$ OpenAlex fields (organized according to the four domains to which they belong), and columns correspond to the incidence growth models: constant incidence (constant), linearly changing incidence (linear), and exponential-growth incidence (exponential). Models that did not converge or for which the parameter p-values were not statistically significant ($P > 0.05$) are indicated with a dash.}
    \label{tbl:fields_}
\end{table}
}

{\footnotesize
\renewcommand{\arraystretch}{0.75}
\setlength{\tabcolsep}{3pt}
\begin{table}[H]
    \centering
    \begin{tabular}{ll|lll}
\textbf{Domain} & \textbf{Field} & $\boldsymbol{\gamma}$ & $\mathbf{g}$ & \textbf{$\mathbf{T_d}$ (years)} \\ \midrule
\multirow{3}{*}{\makecell{Health\\Sciences}} & Dentistry & 0.31 & 0.12 (****) & 5.9 [4.1-10.8] \\
 & Health Professions & 0.09 & 0.14 (****) & 4.9 [4.0-6.2] \\
 & Medicine & 0.02 & 0.06 (****) & 10.8 [9.7-12.1] \\
\midrule
\multirow{3}{*}{\makecell{Life\\Sciences}} & Agricultural and Biological Sciences & 0.10 & 0.11 (****) & 6.2 [5.2-7.5] \\
 & Biochemistry, Genetics and Molecular Biology & 0.10 & 0.14 (****) & 5.0 [4.5-5.7] \\
 & Immunology and Microbiology & 0.01 & 0.06 (****) & 11.7 [9.5-15.0] \\
 & Neuroscience & 0.05 & 0.10 (****) & 7.3 [6.2-8.7] \\
 & Pharmacology, Toxicology and Pharmaceutics & 0.01 & 0.13 (****) & 5.4 [4.5-6.8] \\
 \midrule
\multirow{3}{*}{\makecell{Physical\\Sciences}}& Chemical Engineering & 0.02 & 0.14 (****) & 5.0 [4.0-6.8] \\
 & Chemistry & 0.66 & 0.17 (****) & 4.2 [3.2-6.0] \\
 & Computer Science & 0.46 & 0.24 (****) & 2.8 [2.4-3.5] \\
 & Earth and Planetary Sciences & 0.16 & 0.16 (****) & 4.4 [3.6-5.6] \\
 & Energy & 0.00 & 0.12 (****) & 5.5 [4.6-7.1] \\
 & Engineering & 0.20 & 0.19 (****) & 3.7 [3.2-4.3] \\
 & Environmental Science & 0.12 & 0.14 (****) & 5.1 [4.4-6.2] \\
 & Materials Science & 0.19 & 0.14 (****) & 4.9 [4.1-6.3] \\
 & Mathematics & 0.10 & 0.12 (****) & 6.0 [4.9-7.8] \\
 & Physics and Astronomy & 0.14 & 0.12 (****) & 5.8 [4.8-7.5] \\
 \midrule
\multirow{3}{*}{\makecell{Social\\Sciences}} & Arts and Humanities & 0.12 & 0.11 (****) & 6.3 [4.9-8.8] \\
 & Business, Management and Accounting & 0.85 & 0.15 (****) & 4.5 [3.2-7.4] \\
 & Decision Sciences & 0.39 & 0.16 (****) & 4.2 [3.3-6.1] \\
 & Economics, Econometrics and Finance & 0.49 & 0.15 (****) & 4.6 [3.4-7.1] \\
 & Psychology & 0.13 & 0.09 (****) & 7.9 [6.2-10.7] \\
 & Social Sciences & 0.10 & 0.13 (****) & 5.3 [4.6-6.5] \\
\end{tabular}
    \caption{\textbf{GLM fits of retraction incidence at field-level using an exponential-growth model.}
    Each row corresponds to one field, grouped by their domains. Reported are the estimated dispersion parameter ($\gamma$), the annual growth rate ($g$) with the associated $p$-value, and the doubling time ($T_d$, in years) with its 95\% CI in brackets.
    Significance levels: *~$P~\leq~0.05$; **~$P~\leq~0.01$; ***~$P~\leq~0.001$; ****~$P~\leq~0.0001$.}
    \label{tbl:fields}
\end{table}
}

\section{Retraction Incidence by Country}

We extend our analysis to examine geographic variation in retraction incidence by mapping papers to countries using institutional affiliation data from OpenAlex.

OpenAlex assigns to each work a set of authors, to each author a set of institutional affiliations for that work, and to each institution a country location. To avoid double-counting when works involve multiple authors or when authors have multiple institutional affiliations, we apply fractional weighting that distributes each paper's unit weight proportionally.

Specifically, consider a work $p$ with $A_p$ authors. If author $a$ lists $I_{p,a}$ institutional affiliations in this paper, each spanning potentially different countries, then author $a$ contributes a weight of $1/A_p$ to the work, and this weight is further divided equally among their $I_{p,a}$ affiliations, giving each affiliation a weight of $1/(A_p \cdot I_{p,a})$. The country associated with each affiliation receives this fractional weight. Summing across all authors, a work's total weight of $1$ is distributed across all countries represented in the author-institution-country chains.

For each country and year $t$, we compute the weighted publication total $N_t$ and retraction total $R_t$ by summing these weights across all relevant works. We focus on the $21$ most productive countries in $2021$, which together account for more than $80\%$ of that year's publications (see Fig.~\ref{fig:jitter}B in the main text).

We then apply the same GLM framework to these country-level cohorts, computing BIC weights, dispersion parameters, growth rates, and doubling times for each country.

We initially performed this analysis over the full time window $1992\text{-}2021$. However, the model fits tended to be more influenced by the early period than by the more recent years. To obtain fitted trends that better represent the current retraction landscape, we restrict the country-level analysis to the period $2000\text{-}2021$.

Table~\ref{tbl:countries_} reports the BIC weights for the three incidence models, restricted to the countries that exhibit a strong best-fit model across two time windows: the full period ($1992\text{-}2021$) and the restricted period ($2000\text{-}2021$). 
Notably, all five countries - Australia, China, India, Poland, and the Russian Federation - maintain a BIC weight higher then $91\%$ for exponential growth with positive growth rates.
Table~\ref{tbl:countries} shows the correspondent dispersion parameters, growth rates and doubling times. 
For the full period ($1992\text{-}2021$),  China exhibits the fastest growth with a doubling time of $3.5$ years ($95\%$ CI $[3.0\text{-}4.1]$), followed by the Russian Federation at $4.3$ years ($95\%$ CI $[3.1\text{-}6.8]$), India at $5.3$ years ($95\%$ CI $[4.3\text{-}7.1]$),
Poland at $6.3$ years ($95\%$ CI $[3.8\text{-}16.7]$),
and Australia at $10.0$ years ($95\%$ CI $[7.4\text{-}15.8]$).

These results highlight substantial geographic heterogeneity in retraction-incidence trajectories. While a subset of countries (China, India, and the Russian Federation) exhibits robust exponential growth across both time windows, many others do not show a strong best-fit pattern. 

\begin{figure}[H]
\centering
\centering
\includegraphics[width=1\textwidth]{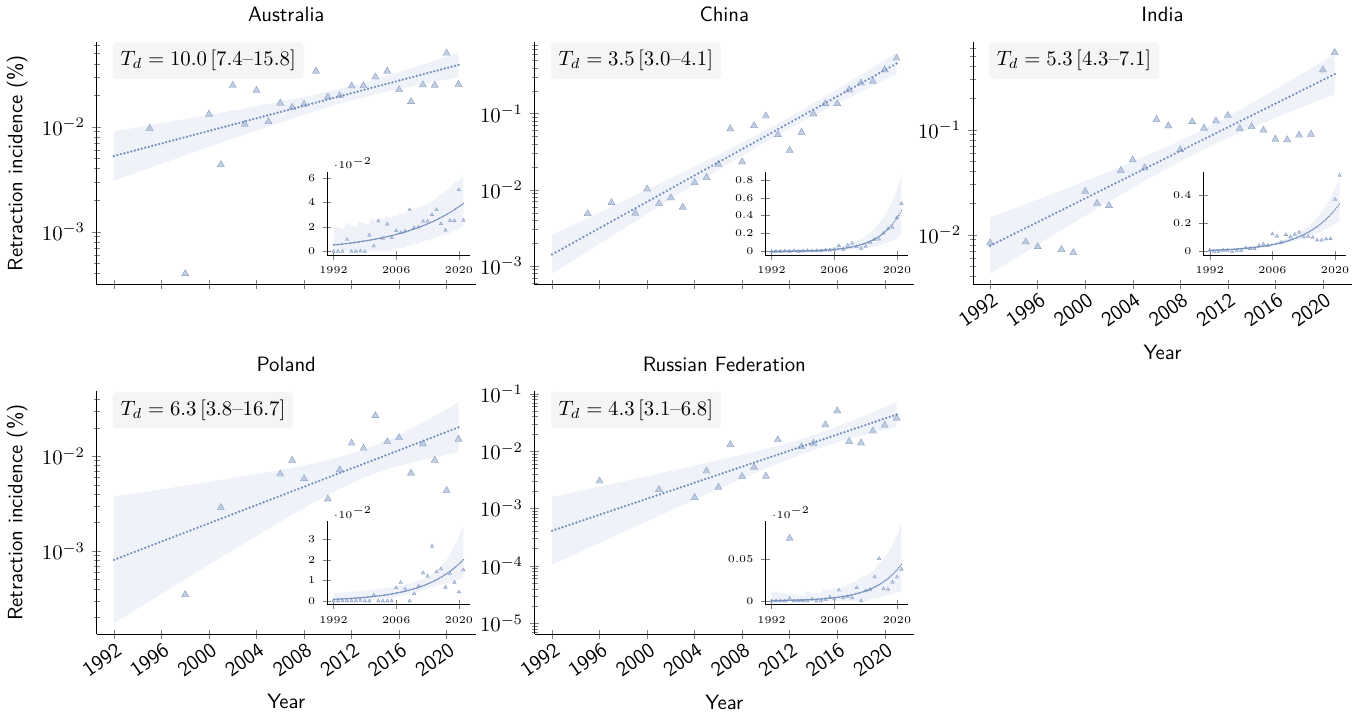}
    \caption{\textbf{
    Retraction incidence over time across countries with a strongly best-fit GLM exponential-growth incidence model.} Each panel representing one of the countries for which the GLM exponential-growth incidence model was strongly best fit (BIC weight $>90\%$)  in both time periods. Blue triangles show the annual retraction incidence (in percentage), and blue lines represent the GLM exponential-growth incidence fits based on a negative binomial likelihood  (
    growth rate significance levels $P \leq 0.05$
    ) with shaded areas indicating the $95\%$ confidence intervals. The doubling times ($T_d$) with their $95\%$ CI are reported in the legend. The $y$-axis is shown on a logarithmic scale (zero values not shown; linear scale with 95\% prediction intervals in inset).}
\label{fig:country exponential}
\end{figure}

\begin{figure}[H]
\centering
\centering
\includegraphics[width=0.8\textwidth]{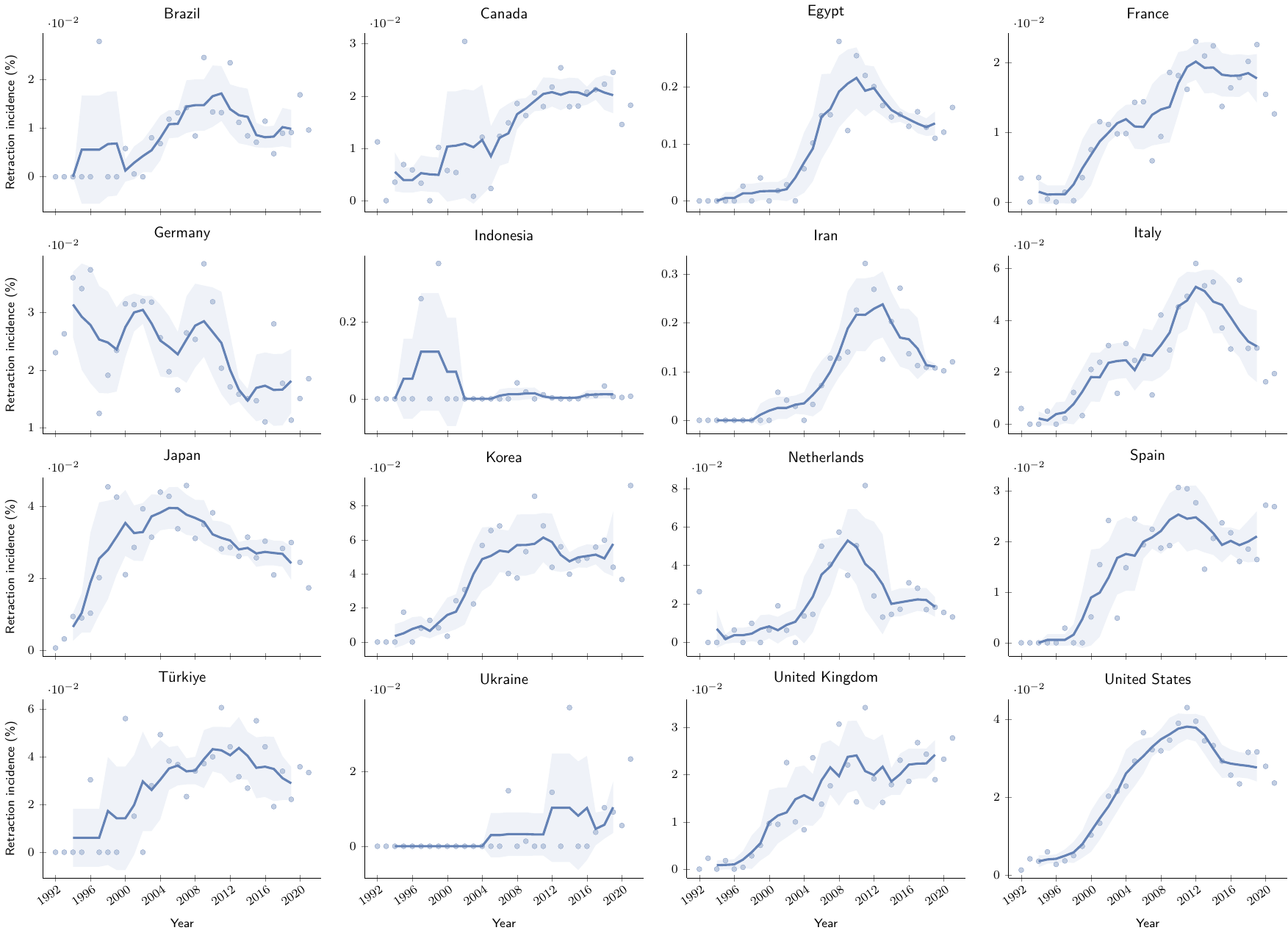}
    \caption{\textbf{
    Retraction incidence over time across countries without a strongly best-fit GLM exponential-growth incidence model.} Each panel represents one of the countries for which no strong exponential-growth fit was observed in both time periods. 
    Blue circles indicate the annual retraction incidence (in percentage), while blue lines show the centered $5$-year rolling average, with shaded areas representing the corresponding standard deviation.
    }
\label{fig:country no exponential}
\end{figure}

Fig.~\ref{fig:country exponential} shows the GLM exponential-growth incidence fits for the $5$ countries where the exponential-growth model achieves strong support (BIC weight $>90\%$ ) in both time periods.
Fig.~\ref{fig:country no exponential} displays retraction incidence patterns for the remaining countries. These countries exhibit diverse temporal dynamics, and several show non-monotonic patterns with peaks followed by recent declines or plateau (e.g., Egypt, France, Iran, Italy, Japan, Korea, Netherlands, Poland, Spain, Türkiye, United States). The recent downward trends in several countries suggest the potential impact of policy interventions and improved research integrity workflows at the country level. 

{\footnotesize
\renewcommand{\arraystretch}{0.75}
\setlength{\tabcolsep}{3pt}
\begin{table}[H]
    \centering
    \begin{tabular}{l|ccc|ccc}
 & \multicolumn{3}{c}{1992--2021} & \multicolumn{3}{|c}{2000--2021} \\
\midrule
\textbf{Country/Fit} & \textbf{Constant} & \textbf{Linear} & \textbf{Exponential} & \textbf{Constant} & \textbf{Linear} & \textbf{Exponential} \\
\midrule
Australia & 0.0\% & - & \textbf{100.0\%} & 2.9\% & 0.4\% & \textbf{96.7\%} \\
China & 0.0\% & - & \textbf{100.0\%} & 0.0\% & - & \textbf{100.0\%} \\
India & 0.0\% & - & \textbf{100.0\%} & 0.0\% & 9.0\% & \textbf{91.0\%} \\
Poland & 0.2\% & - & \textbf{99.8\%} & 6.3\% & - & \textbf{93.7\%} \\
Russian Federation & 0.0\% & - & \textbf{100.0\%} & 0.0\% & - & \textbf{100.0\%} \\
\end{tabular}
    \caption{
    \textbf{GLM fits of retraction incidence at the country level for countries showing a strong best-fit.} The table shows the BIC weights in percentage, with the highest weight highlighted in bold. Rows represent the three countries that exhibited a strong best-fit pattern across both time windows, and columns correspond to the incidence-growth models: constant incidence (constant), linearly changing incidence (linear), and exponential-growth incidence (exponential). Results are shown for two time windows: $1992\text{-}2021$ (left) and $2000\text{-}2021$ (right). Models that did not converge or for which the parameter p-values were not statistically significant ($P > 0.05$) are indicated with a dash.
    }
    \label{tbl:countries_}
\end{table}
}

{\footnotesize
\renewcommand{\arraystretch}{0.75}
\setlength{\tabcolsep}{3pt}
\begin{table}[H]
    \centering
    \begin{tabular}{l|lll|lll}
 & \multicolumn{3}{c}{1992--2021} & \multicolumn{3}{|c}{2000--2021} \\
\midrule
\textbf{Country} & $\boldsymbol{\gamma}$ & $\mathbf{g}$ & \textbf{$\mathbf{T_d}$ (years)} & $\boldsymbol{\gamma}$ & $\mathbf{g}$ & \textbf{$\mathbf{T_d}$ (years)} \\
\midrule
Australia & 0.04 & 0.07 (****) & 10.0 [7.4-15.8] & 0.02 & 0.04 (***) & 16.6 [10.2-45.1] \\
China & 0.14 & 0.20 (****) & 3.5 [3.0-4.1] & 0.14 & 0.20 (****) & 3.6 [3.1-4.2] \\
India & 0.26 & 0.13 (****) & 5.3 [4.3-7.1] & 0.22 & 0.10 (****) & 7.0 [5.1-11.1] \\
Poland & 0.23 & 0.11 (**) & 6.3 [3.8-16.7] & 0.19 & 0.09 (**) & 7.7 [4.2-39.9] \\
Russian Federation & 0.21 & 0.16 (****) & 4.3 [3.1-6.8] & 0.21 & 0.15 (****) & 4.5 [3.2-7.8] \\
\end{tabular}
    \caption{
    \textbf{GLM fits of retraction incidence at country-level using an exponential-growth model.}
    Each row corresponds to one country. Reported are the estimated dispersion parameter ($\gamma$), the annual growth rate ($g$) with the associated $p$-value, and the doubling time ($T_d$, in years) with its 95\% CI in brackets. Results are shown for two time windows: $1992\text{-}2021$ (left) and $2000\text{-}2021$ (right). Only countries with strong exponential fits ($100\%$ BIC weight) are reported for each period. Significance levels: *~$P~\leq~0.05$; **~$P~\leq~0.01$; ***~$P~\leq~0.001$; ****~$P~\leq~0.0001$.
    }
    \label{tbl:countries}
\end{table}
}

\section{Retraction Incidence by Publisher}
We examine retraction incidence patterns across publishers by assigning each work to its journal and aggregating at the publisher level. Each journal in OpenAlex is associated with a specific publisher, allowing us to group works by their publishing organization.n We focus on the top $20$ publishers by publication volume in $2021$, which represent the major players in academic publishing.
Note that some publishers were founded after $1992$, the start of the period under analysis. In those cases, we restrict our analysis to begin from their year of foundation. Specifically: Multidisciplinary Digital Publishing Institute ($1996$), Hindawi Publishing Corporation ($1997$), Medknow ($1997$), Lippincott Williams \& Wilkins ($1998$), BioMed Central ($2000$), and Frontiers Media ($2007$).

Table~\ref{tbl:publishers_} shows BIC weights comparing the three incidence models across publishers. The exponential-growth model provides the best fit for $18$ out of $20$ publishers, achieving $100\%$ BIC weight in $14$ of them. This suggests that exponential growth in retraction incidence characterizes the majority of the publishers. However, some exceptions emerge. Lippincott Williams \& Wilkins and Medknow show $100\%$ BIC weight for the constant-incidence model, suggesting stable retraction rates over time. Several publishers show mixed evidence: Nature Portfolio ($77.6\%$ exponential vs $22.4\%$ constant), and Multidisciplinary Digital Publishing Institute ($92.5\%$ exponential vs $7.5\%$ constant).

Table~\ref{tbl:publishers} reports dispersion parameters, growth rates and doubling times for the $16$ publishers where the exponential model achieves strong support. 
IOP Publishing ($2.7$ years, $95\%$ CI $[2.2\text{-}3.7]$)), and 
Hindawi Publishing Corporation ($3.3$ years, $95\%$ CI $[2.5\text{-}4.9]$)exhibits the fastest growth, followed closely by 
Royal Society of Chemistry at $3.6$ years ($95\%$ CI $[2.7\text{-}5.3]$), 
EDP Sciences at $3.7$ years ($95\%$ CI $[2.2\text{-}11.1]$), 
SAGE Publishing at $3.7$ years ($95\%$ CI $[3.2\text{-}4.6]$), Institute of Electrical and Electronics Engineers at $3.8$ years ($95\%$ CI $[2.5\text{-}7.5]$),
and Cambridge University Press ($3.8$ years, $95\%$ CI $[2.7\text{-}6.5]$) 
Several major publishers show intermediate growth rates: 
Taylor \& Francis ($4.4$ years, $95\%$ CI $[3.8\text{-}5.1]$),
Springer Science+Business Media ($4.9$ years, $95\%$ CI $[4.3\text{-}5.6]$),
and Wiley ($5.4$ years, $95\%$ CI $[4.7\text{-}6.4]$).
Oxford University Press, The American Chemical Society, BioMed Central, Springer Nature, and Frontiers Media show slower growth with doubling times ranging from $6.0$ to $8.1$ years. 

\begin{figure}[H]
\centering
\includegraphics[width=\textwidth, height=0.85\textheight, keepaspectratio]{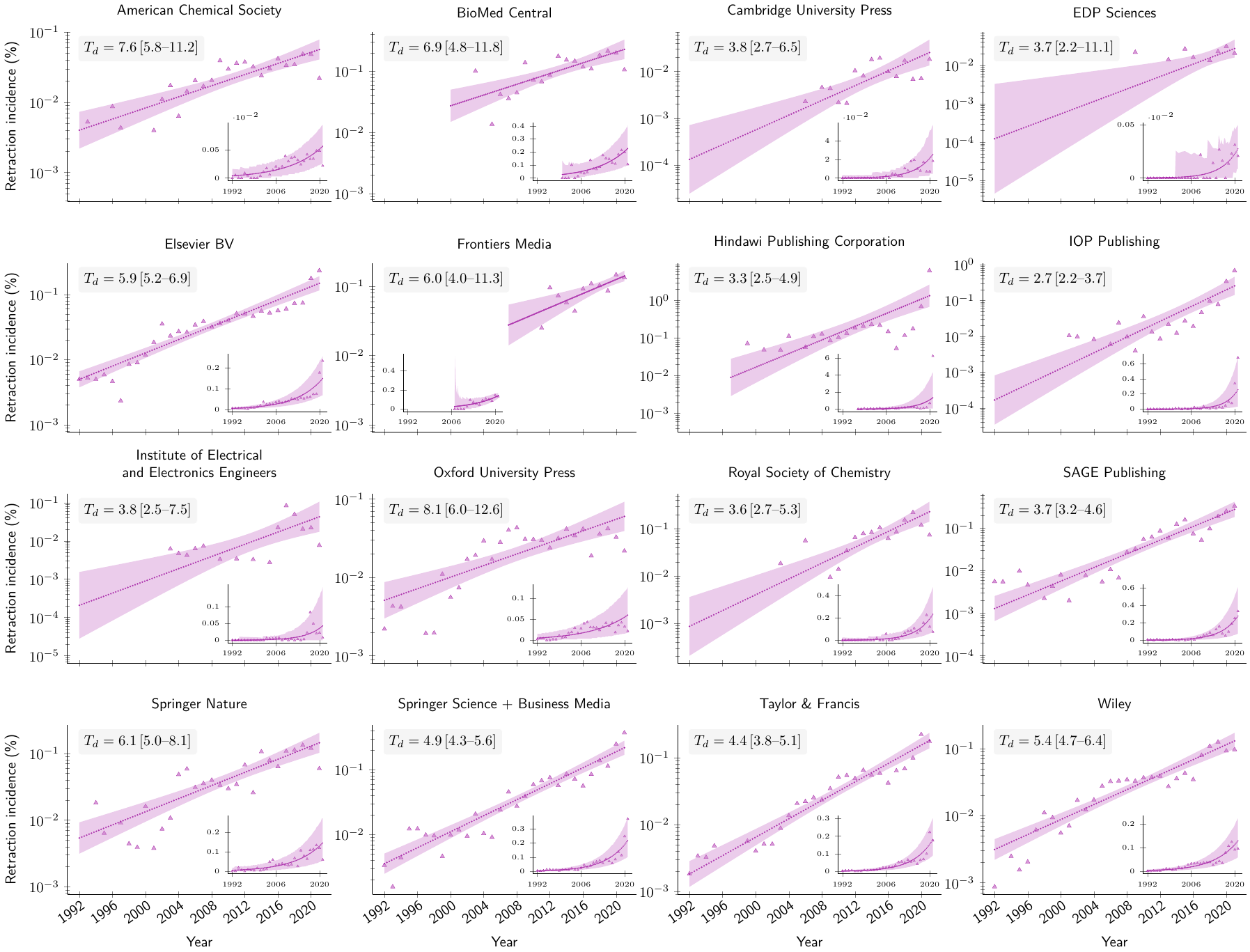}
    \caption{\textbf{
    Retraction incidence over time across publishers with a strongly best-fit GLM exponential-growth incidence model.} Each panel representing one of the publishers for which the GLM exponential-growth incidence model was strongly best fit (BIC weight $>90\%$). Magenta triangles show the annual retraction incidence (in percentage), and magenta lines represent the GLM exponential-growth incidence fits based on a negative binomial likelihood, with shaded areas indicating the $95\%$ confidence intervals. The doubling times ($T_d$) with their $95\%$ CI are reported in the legend. Growth rate significance levels $P \leq 0.05$.
    The $y$-axis is shown on a logarithmic scale (zero values not shown; linear scale with 95\% prediction intervals in inset).
    }
\label{fig:publisher exponential}
\end{figure}

\begin{figure}[H]
\centering
\includegraphics[width=\textwidth, height=1\textheight, keepaspectratio]{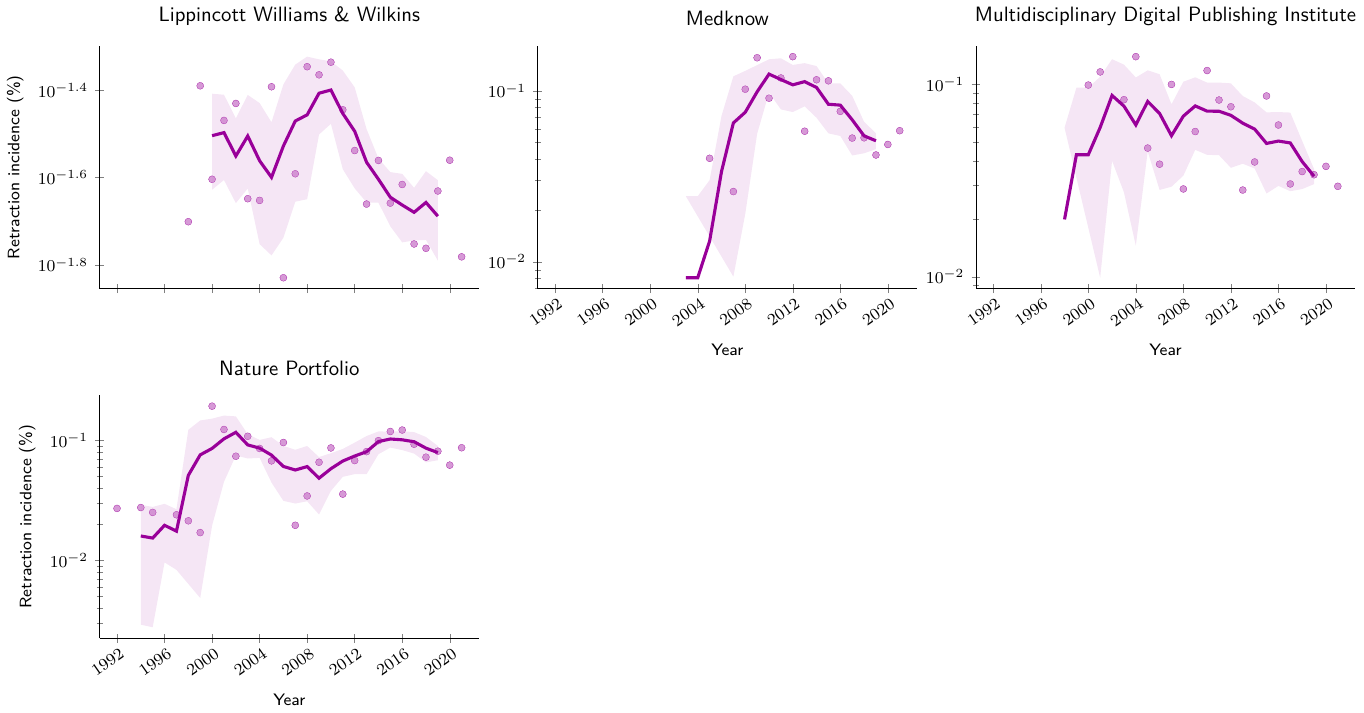}
    \caption{\textbf{
    Retraction incidence over time across publishers without a strongly best-fit GLM exponential-growth incidence model.} Each panel represents one of the publishers for which no strong exponential-growth fit was observed. 
    Magenta circles indicate the annual retraction incidence (in percentage), while magenta lines show the centered $5$-year rolling average, with shaded areas representing the corresponding standard deviation.
    }
\label{fig:publisher no exponential}
\end{figure}

Figure~\ref{fig:publisher exponential} shows the GLM exponential-growth incidence fits for the $16$ publishers where the exponential-growth model achieves strong support (BIC weight $>90\%$ ).
Figure~\ref{fig:publisher no exponential} displays retraction incidence patterns for the remaining publishers. 

{\footnotesize
\renewcommand{\arraystretch}{0.75}
\setlength{\tabcolsep}{3pt}
\begin{table}[H]
    \centering
    \begin{tabular}{l|ccc}
\textbf{Publisher/Fit} & \textbf{Constant} & \textbf{Linear} & \textbf{Exponential} \\ \midrule
American Chemical Society & 0.0\% & - & \textbf{100.0\%} \\
BioMed Central & 0.0\% & - & \textbf{100.0\%} \\
Cambridge University Press & 0.0\% & - & \textbf{100.0\%} \\
EDP Sciences & 0.1\% & - & \textbf{99.9\%} \\
Elsevier BV & 0.0\% & 0.0\% & \textbf{100.0\%} \\
Frontiers Media & 0.1\% & - & \textbf{99.9\%} \\
Hindawi Publishing Corporation & 0.0\% & - & \textbf{100.0\%} \\
IOP Publishing & 0.0\% & - & \textbf{100.0\%} \\
Institute of Electrical and Electronics Engineers & 0.0\% & - & \textbf{100.0\%} \\
Lippincott Williams \& Wilkins & \textbf{100.0\%} & 0.0\% & - \\
Medknow & \textbf{100.0\%} & - & - \\
Multidisciplinary Digital Publishing Institute & 7.5\% & - & \textbf{92.5\%} \\
Nature Portfolio & 22.4\% & - & \textbf{77.6\%} \\
Oxford University Press & 0.0\% & - & \textbf{100.0\%} \\
Royal Society of Chemistry & 0.0\% & - & \textbf{100.0\%} \\
SAGE Publishing & 0.0\% & - & \textbf{100.0\%} \\
Springer Nature & 0.0\% & - & \textbf{100.0\%} \\
Springer Science+Business Media & 0.0\% & - & \textbf{100.0\%} \\
Taylor \& Francis & 0.0\% & - & \textbf{100.0\%} \\
Wiley & 0.0\% & - & \textbf{100.0\%} \\
\end{tabular}
    \caption{\textbf{GLM fits of retraction incidence at publisher-level under different growth models.} The table shows the BIC weights in percentage, with the highest weight highlighted in bold. Rows correspond to the $20$ publishers under analysis (in alphabetical order), and columns correspond to the incidence growth models: constant incidence (constant), linearly changing incidence (linear), and exponential-growth incidence (exponential). Models that did not converge or for which the parameter p-values were not statistically significant ($P > 0.05$) are indicated with a dash.}
    \label{tbl:publishers_} 
\end{table}
}

{\footnotesize
\renewcommand{\arraystretch}{0.75}
\setlength{\tabcolsep}{3pt}
\begin{table}[H]
    \centering
    \begin{tabular}{l|lll}
\textbf{Publisher} & $\boldsymbol{\gamma}$ & $\mathbf{g}$ & \textbf{$\mathbf{T_d}$ (years)} \\ \midrule
American Chemical Society & 0.07 & 0.09 (****) & 7.6 [5.8-11.2] \\
BioMed Central & 0.11 & 0.10 (****) & 6.9 [4.8-11.8] \\
Cambridge University Press & 0.20 & 0.18 (****) & 3.8 [2.7-6.5] \\
EDP Sciences & 0.00 & 0.19 (**) & 3.7 [2.2-11.1] \\
Elsevier BV & 0.10 & 0.12 (****) & 5.9 [5.2-6.9] \\
Frontiers Media & 0.00 & 0.12 (****) & 6.0 [4.0-11.3] \\
Hindawi Publishing Corporation & 0.72 & 0.21 (****) & 3.3 [2.5-4.9] \\
IOP Publishing & 0.46 & 0.25 (****) & 2.7 [2.2-3.7] \\
Institute of Electrical and Electronics Engineers & 0.97 & 0.18 (****) & 3.8 [2.5-7.5] \\
Oxford University Press & 0.20 & 0.09 (****) & 8.1 [6.0-12.6] \\
Royal Society of Chemistry & 0.19 & 0.19 (****) & 3.6 [2.7-5.3] \\
SAGE Publishing & 0.25 & 0.19 (****) & 3.7 [3.2-4.6] \\
Springer Nature & 0.14 & 0.11 (****) & 6.1 [4.9-8.1] \\
Springer Science+Business Media & 0.10 & 0.14 (****) & 4.9 [4.3-5.6] \\
Taylor \& Francis & 0.09 & 0.16 (****) & 4.4 [3.8-5.1] \\
Wiley & 0.11 & 0.13 (****) & 5.4 [4.7-6.4] \\
\end{tabular}
    \caption{\textbf{GLM fits of retraction incidence at publisher-level using an exponential-growth model.}
    Each row corresponds to one publisher. Reported are the estimated dispersion parameter ($\gamma$), the annual growth rate ($g$) with the associated $p$-value, and the doubling time ($T_d$, in years) with its 95\% CI in brackets.
    Significance levels: *~$P~\leq~0.05$; **~$P~\leq~0.01$; ***~$P~\leq~0.001$; ****~$P~\leq~0.0001$.}
    \label{tbl:publishers}
\end{table}
}

\section{Country and Publisher-level coverage of total and retracted publications}

In this section, we provide detailed analysis of country-level concentration patterns for both total and retracted publications, complementing the findings presented in the main text.


\begin{figure}[H]
\centering
\includegraphics[width=\textwidth, height=0.85\textheight, keepaspectratio]{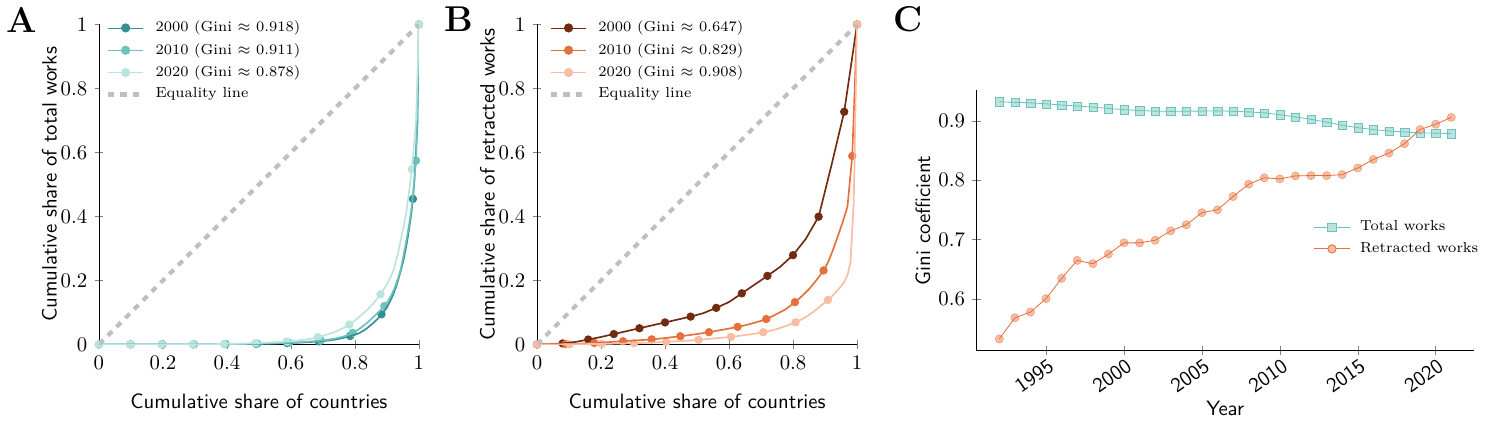}
\caption{\textbf{Country-level Lorenz curves and Gini coefficients of total and retracted publications.} A-B. Lorenz curves of papers across countries at ten-year intervals (2000–2020). Corresponding Gini coefficients are shown in the legend. Equality line shown as dotted diagonal. A: total publications; B: total retractions. C. Annual country-level Gini coefficient for total and retracted publications (1992–2021, 5-year centered rolling average).}
\label{fig:Gini_countries}
\end{figure}

As shown in the main text (Figure \ref{fig:jitter}B-C), we observe a striking divergence in geographic concentration over time. While the number of countries needed to account for 90\% of total publications increases steadily from 1992 to 2021, indicating greater global distribution of scientific output, the number of countries accounting for 90\% of retractions shows the opposite trend, suggesting increasing concentration of integrity issues in fewer countries.
To further examine this divergence, we employ Lorenz curves and Gini coefficients as complementary measures of inequality. Figure~\ref{fig:Gini_countries} presents Lorenz curves for both total publications (Panel A) and retractions (Panel B) at ten-year intervals (2000, 2010, 2020). For total publications, the curves move closer to the equality line over time, reflecting decreasing concentration. In contrast, the Lorenz curves for retractions move away from the equality line, indicating increasing concentration among fewer countries.
Figure~\ref{fig:Gini_countries} shows the temporal evolution of Gini coefficients from 1992 to 2021. The Gini coefficient for total works exhibits a consistent downward trend (from approximately 0.92 to 0.88), confirming the democratization of scientific publishing. Conversely, the Gini coefficient for retracted works shows an upward trajectory (from approximately 0.55 to 0.90), reinforcing the pattern of increasing concentration observed in the main text analysis.
For all calculations, we restrict the analysis to countries with at least one publication (for total works) and to countries with at least one retraction (for retracted works). 

\vspace{10pt}
\begin{figure}[H]
    \centering
    \begin{subfigure}[b]{0.45\textwidth}
        \centering
        \begin{overpic}[width=\textwidth]{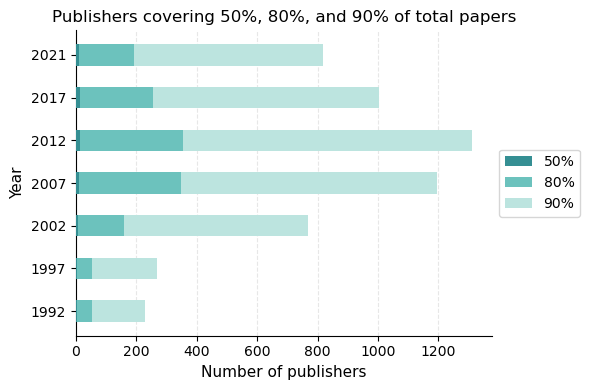}
            \put(0,70){\textbf{A}} 
        \end{overpic}
    \end{subfigure}
    \hfill
    \begin{subfigure}[b]{0.45\textwidth}
        \centering
        \begin{overpic}[width=\textwidth]{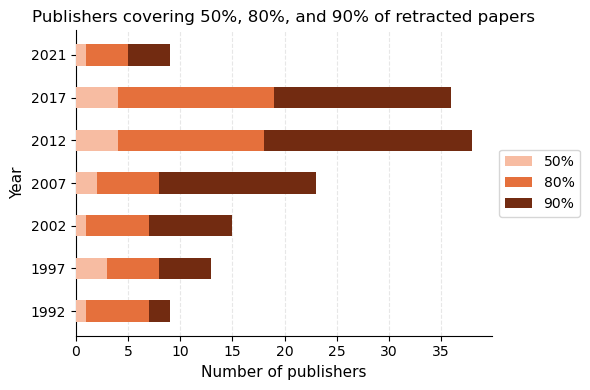}
            \put(0,70){\textbf{B}} 
        \end{overpic}
    \end{subfigure}
        \caption{\textbf{Publisher-level coverage of total and retracted publications.}
    Number of publishers required to account for fixed global shares of output at five-year intervals (1992–2021). A: total publications; B: total retractions. Bars report the publisher counts needed to reach 50\%, 80\%, and 90\% of the respective totals at each time point.}
    \label{fig:Coverage_publishers}
\end{figure}

We extend the country-level analysis to examine concentration patterns at the publisher level. Figure~\ref{fig:Coverage_publishers} shows the number of publishers required to account for 50\%, 80\%, and 90\% of total publications (Panel A) and retractions (Panel B) at five-year intervals from 1992 to 2021.
For total publications, we observe substantial growth in the number of publishers needed to reach each threshold. The number of publishers accounting for 90\% of output increased from approximately 200 in 1992 to over 1,300 in 2012, before declining to around 800 in 2021.
On the other side, the number of publishers accounting for 90\% of retractions ranges from approximately 10-35 across the study period, representing extreme concentration. Notably, in 2021, just 2-3 publishers account for 50\% of all retractions, and fewer than 10 publishers account for 80\%.
This publisher-level concentration of retractions appears even more pronounced than the country-level concentration, suggesting that integrity issues may be concentrated not only geographically but also within specific publishing venues. 


\begin{figure}[H]
\centering
\includegraphics[width=\textwidth, height=0.85\textheight, keepaspectratio]{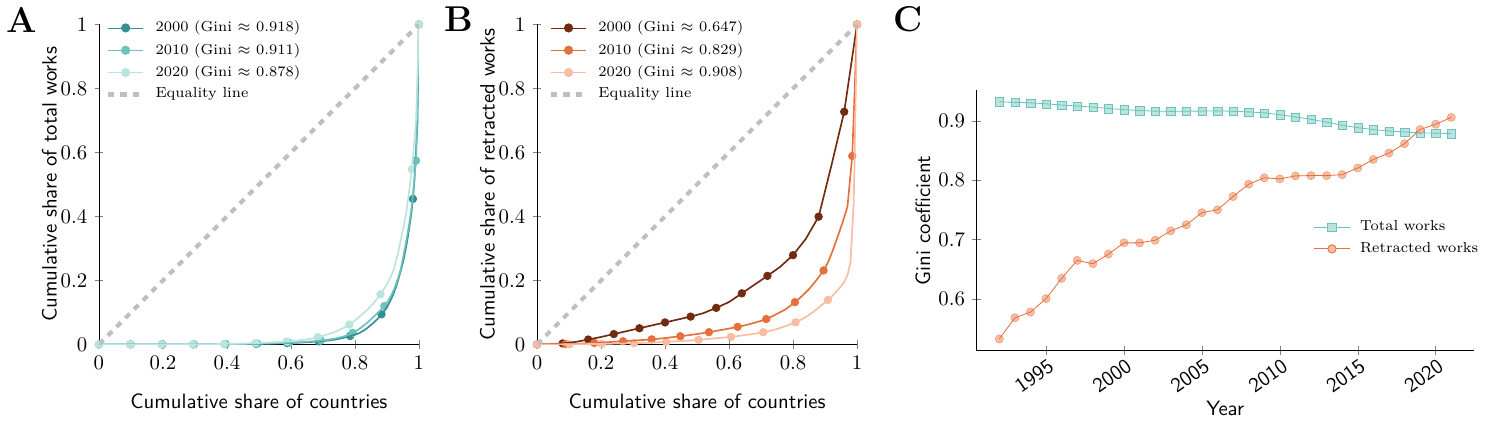}
\caption{\textbf{Publisher-level Lorenz curve and Gini coefficient of total and retracted publications.} A-B. Lorenz curves of papers across publishers at ten-year inter-
vals (2000–2020). Corresponding Gini coefficients are shown in the legend. Equality line shown as dotted diagonal. A: total publications; B: total retractions. C. Annual publisher-level Gini coefficient for total and retracted publications (1992–2021, 5-year centered rolling average).}
\label{fig:Gini_publishers}
\end{figure}

In Figure~\ref{fig:Gini_publishers}, publisher-level inequality measures are reported. 
For all calculations, we restrict the analysis to publishers with at least one publication (for total works) and to publishers with at least one retraction (for retracted works). 
Panel A shows that total publications concentration stay stable across 2000-2020 (Gini coefficient around 0.90). Panel B reveals that retractions exhibit a Gini coefficients rising from 0.576 (2000) to 0.886 (2020).
Panel C confirms these patterns over the full time period. Total works maintain a constant concentration (Gini around 0.90-0.92), while retracted works show an increases from Gini 0.45 to 0.90.

\section{Abbreviations}
Table~\ref{tbl:abbreviations}  provides the fields and publisher abbreviations displayed in Fig.~\ref{fig:domains}B and Fig.~\ref{fig:jitter}A in the main text.

{\footnotesize
\renewcommand{\arraystretch}{0.75}
\setlength{\tabcolsep}{3pt}
\begin{table}[H]
    \centering
    \begin{tabular}{l | l l}
 &\textbf{Name} & \textbf{Abbreviation} \\ \midrule
Fields & Health Professions & Health Prof.\\
 & Pharmacology, Toxicology and Pharmaceutics & Pharm. Tox. Pharm.\\
 &Biochemistry, Genetics and Molecular Biology & Bio. Gen. Mol. Biol.\\
 &Agricultural and Biological Sciences & Agri. Biol. Sci.\\
 &Immunology and Microbiology & Imm. Micro.\\
 &Computer Science & Comp. Sci.\\
 &Earth and Planetary Sciences & Earth Planet. Sci.\\
 &Materials Science & Materials Sci.\\
 &Chemical Engineering & Chemical Eng.\\
 &Physics and Astronomy & Phys. Astron.\\
 &Environmental Science & Environ. Sci.\\
 &Decision Sciences & Decision Sci.\\
 &Economics, Econometrics and Finance & Econ. Econ. Fin.\\
 &Business, Management and Accounting & Bus. Manag. Acc.\\
 &Social Sciences & Soc. Sci.\\
 &Arts and Humanities & Arts Hum.\\
\midrule
Publisher &Hindawi Publishing Corporation & Hindawi\\
 &IOP Publishing & IOP\\
 &Multidisciplinary Digital Publishing Institute & MDPI\\
 &Springer Nature & Springer\\
 &Elsevier BV & Elsevier\\
\end{tabular}
    \caption{\textbf{Fields and publisher abbreviations.} The table shows fields and publisher abbreviations displayed in Fig.~\ref{fig:domains}B and Fig.~\ref{fig:jitter}A in the main text.}
    \label{tbl:abbreviations}
\end{table}
}

\end{document}